\documentclass[12pt]{article}
\usepackage{latexsym,,amssymb,amsfonts}
\usepackage[dvips]{graphicx}
\begin{document}
\begin{titlepage}
\begin{flushright}
\par\vglue -2cm
CERN--TH 98-351\\
ISN 98--21
\end{flushright}
\vfill
\begin{center}
{\Large{\bf ON~THE~STABILITY~DOMAIN~OF~SYSTEMS}}\\
\vskip .4cm
{\Large{\bf OF THREE  ARBITRARY CHARGES$^\dagger$}}\\
\vskip .6cm 
{\bf  Ali Krikeb}$^{(1)}$, {\bf Andr\'e Martin}$^{(2), (3)}$,\\[.4cm] 
{\bf Jean-Marc Richard}$^{(4)}$ and  
{\bf Tai Tsun Wu}$^{\,\star}$$\,^{(2),(5)}$\\
\vskip .5cm
{\small $^{(1)}\,$ Institut de Physique Nucl\'eaire de Lyon-CNRS-IN2P3}\\[-.05cm]
{\small Universit\'e Claude Bernard, Villeurbanne, France}
\vskip .3cm
 {\small $^{(2)}\,$ CERN, Theory Division, CH 1211 Gen\`eve 23}\\
\vskip .3cm
{\small $^{(3)}\,$LAPP,  B.P.~110, F-74941 Annecy Le Vieux, France}\\
\vskip .3cm
{\small $^{(4)}\,$Institut des Sciences Nucl\'eaires--CNRS--IN2P3}\\[-.05cm]
{\small Universit\'e Joseph Fourier}\\[-.05cm]
{\small 53, avenue des Martyrs, F-38026 Grenoble, France}\\
\vskip .3cm
{\small $^{(5)}\,$Gordon McKay Laboratory}\\[-.05cm]
{\small Harvard University, Cambridge, Massachusetts}\\
\vglue 1.5cm
{\bf Abstract}\\
\vglue.15cm
\end{center}
We present results on the stability of quantum systems consisting of a 
negative charge $-q_1$ with mass $m_{1}$ and two positive charges $q_2$ 
and $q_3$, with masses $m_{2}$ and $m_{3}$, respectively.  We show that, 
for given masses $m_{i}$, each instability domain is convex in the 
plane of the variables $(q_{1}/q_{2},\,q_{1}/q_{3})$.  A new proof is 
given of the instability of muonic ions $(\alpha, p, \mu^-)$.  We 
then study stability  in some critical regimes where $q_3\ll 
q_2$: stability is sometimes restricted to large values of some mass 
ratios;  the behaviour of the stability frontier is established to 
leading order in $q_3/q_2$.  Finally we present some conjectures about 
the shape of the stability domain, both for given masses and varying 
charges, and for given charges and varying masses.\par
\vfill
\begin{flushleft}
{\it $^{\dagger}\,$ Dedicated to the memory of Harry Lehmann}\\[.1cm]
{\small $^{\star}\,$Work supported in part by the U.S.~Department of Energy under
Grant No.~DE-FG02-84-ER40158.}\\[.2cm]
CERN--TH 98-351\\
\today.
\end{flushleft}
\end{titlepage}

\section{Introduction}
\label{se:Introduction}
In two previous papers \cite{MRW,MRW-general}, hereafter referred to 
as I and II, respectively, we studied the stability of quantum 
systems consisting of point-like electric charges 
\begin{equation}
\label{def-charges}
Q_{i}=\pm[-q_1,q_{2},q_{3}],\qquad q_{i}> 0,
\end{equation}
and masses $m_i$. The Hamiltonian is
\begin{equation}
	H={\vec{\rm p}_{1}^{2}\over 2 m_{1}}+{\vec{\rm p}_{2}^{2}\over 2 m_{2}}
	+{\vec{\rm p}_{3}^{2}\over 2 m_{3}}-{q_{12}\over r_{12}}
	-{q_{13}\over r_{13}}+{q_{23}\over r_{23}},
	\label{Hamiltonian}
\end{equation}
where $q_{ij}=q_{i}q_{j}$ and $r_{ij}=\vert \vec{\rm r}_{i}-\vec{\rm 
r}_{j}\vert$.

In I, we considered the case of unit charges $q_i=1$, with 
application to physical systems such as ${\rm H}{_2}^+({\rm e}^{-}{\rm 
p}{\rm p})$, ${\rm H}^{-}({\rm p}{\rm e}^{-}{\rm e}^{-})$ and ${\rm 
Ps}^{-} ({\rm e}^+{\rm e}^{-}{\rm e}^{-})$, which are stable, or 
$({\rm e}^{-}{\rm p}{\rm e}^{+})$ which is unbound.  We pointed out 
simple properties of the stability domain, leading to a unified 
presentation of results already known \cite{Armour93}, and to a number 
of new results.  For instance, if $m$ is the largest proton mass which gives 
a stable $(e^{-},e^{+},p)$ system when associated to a positron and an 
electron both of mass $m_{e}=1$, it is found in I that $m<4.2$, a 
significant improvement over previous bounds \cite{Armour93}.  This 
means that global considerations on the stability domain can sometimes 
complement specific studies adapted to particular mass configurations.

In II, we extended the discussion by letting the charges $q_i$ themselves
vary. The number of parameters is increased from two to four, and one can 
choose two mass ratios and two charge ratios. The general properties of 
the stability domain established in II  will be briefly reviewed, and 
supplemented by new results. This will be the subject of 
Sec.~\ref{se:Gen-resu}.

As an example of application of general considerations on the 
stability domain, we shall present in Sec.~\ref{se:Alpha} a new proof 
that muonic ions involving a  helium nucleus $\alpha$, such as 
$(p,\alpha,\mu^-)$ or $(d,\alpha,\mu^-)$, are not stable.  This 
confirms results \cite{Chen90} obtained previously using the 
Born--Oppenheimer framework.

In II, we also considered the limiting case where 
$q_{3}\rightarrow0$, but only in the Born--Oppenheimer case where 
$m_{2}=m_{3}=\infty$. We shall resume in Sec.~\ref{se:small-q3} our 
investigations and study the behaviour of the stability frontier in the 
case where $q_{2}\gtrsim1$ and $q_{3}\ll q_{2}$.

In Sec.~\ref{se:Outlook}, we shall present some speculations about 
the plausible shape of the stability domain in both representations: 
varying charge-ratios for given masses, or varying masses for given 
charges. A number of interesting questions remain open.

Our rigorous results are supplemented by numerical investigations 
based on a variational approximation to the solution of the 3-body 
Schr{\"o}dinger equation. In particular, we display an estimate of 
the domain of stability in the $(q_{2},\,q_{3})$ plane for some given 
sets of constituent masses.

\section{General properties of the stability domain}
\label{se:Gen-resu}
\subsection{Inverse-mass plane for unit charges}
\label{sub:Gen-resu:Unit}
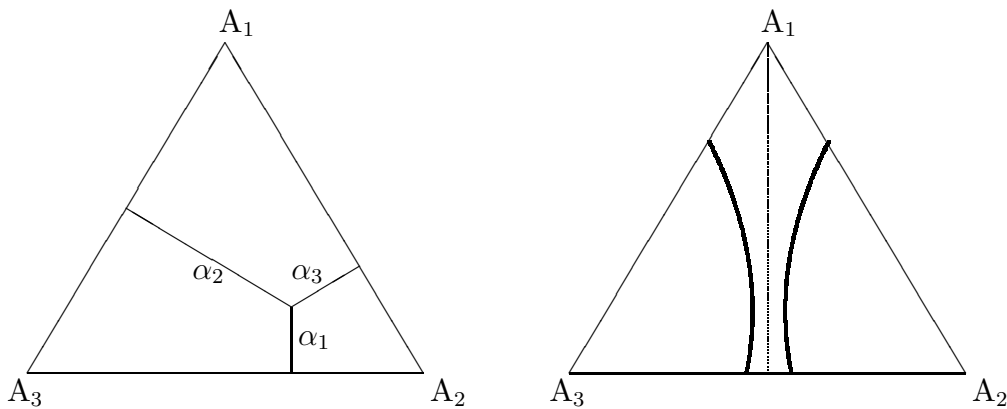
\begin{figure}[htbc]
\setbox1=\vbox{
 \setlength{\unitlength}{0.25pt}
  \begin{picture}(1200,700)(0,0)
  \font\gnuplot=cmr10 at 12pt
  \gnuplot
  \put(300,100){\line(1,0){600}}
  \put(300,100){\line(3,5){300}}
  \put(900,100){\line(-3,5){300}}
  \put(590,630){\makebox(0,0)[l]{A$_1$}}
  \put(910,70){\makebox(0,0)[l]{A$_2$}}
  \put(270,70){\makebox(0,0)[l]{A$_3$}}
  \put(700,200){\line(0,-1){100}}
  \put(700,200){\line(-5,3){250}}
  \put(700,200){\line(5,3){103}}
  \put(710,150){\makebox(0,0)[l]{$\alpha_1$}}
  \put(550,250){\makebox(0,0)[l]{$\alpha_2$}}
  \put(700,250){\makebox(0,0)[l]{$\alpha_3$}}
   \end{picture}
  }
\setbox2=\vbox{
 \setlength{\unitlength}{0.25pt}
  \begin{picture}(1200,700)(0,0)
  \font\gnuplot=cmr10 at 12pt
  \gnuplot
  \linethickness{.7pt}
  \put(300,100){\line(1,0){600}}
  \put(300,100){\line(3,5){300}}
  \put(900,100){\line(-3,5){300}}
   \linethickness{.4pt}
  \qbezier[120](600,100)(600,400)(600,600)
  \put(590,630){\makebox(0,0)[l]{A$_1$}}
  \put(910,70){\makebox(0,0)[l]{A$_2$}}
  \put(270,70){\makebox(0,0)[l]{A$_3$}}
\linethickness{1.2pt}
\qbezier(510,450)(600,270)(566,100)
\qbezier(690,450)(600,270)(634,100)
  \end{picture}
 }
\centerline{\hspace{3.5cm}\box1\hspace{ -6.5cm}\box2\hfil}
 \caption{
\label{Fig1} 
Domain of possible inverse masses $\alpha_i$, normalised by 
$\sum\alpha_i=1$, and shape of the stability domain for three unit 
charges. }
\end{figure}

Consider first the case where $q_{1}=q_{2}=q_{3}$, which can be 
chosen as $q_{i}=1$. Stability is 
defined as the existence of a normalised 3-body bound state with an 
energy below that of the lowest (1,2) or (1,3) atom, i.e.,
\begin{equation}
	E^{(3)}<\min\left(E^{(2)}_{12},E^{(2)}_{13}\right),\qquad
	E^{(2)}_{1i}=-(\alpha_{1}+\alpha_{i})^{-1}/2,
	\label{def-stab}
\end{equation}
where $\alpha_{i}=1/m_{i}$ is the inverse mass of particle $i$. Thanks 
to scaling, there are only two independent mass ratios in this problem.
In I, we found it convenient to represent any possible system as a 
point inside the triangle of inverse masses normalised by 
$\sum_{i=1}^{3}\alpha_{i}=1$. This triangular plot is shown in Fig.\ \ref{Fig1}.

In this representation, the stability domain appears as  a band around 
the symmetry axis where $\alpha_{2}=\alpha_{3}$. It is schematically 
pictured in Fig.\ \ref{Fig1}.

The following rigorous properties are known, or shown in I.

\noindent{\sl a)} All points of the symmetry axis ($\alpha_{2}=\alpha_{3}$, 
$0\le\alpha_{1}\le 1$) belong  to the stability domain \cite{Hill77}.

\noindent{\sl b)} Each instability domain is star shaped with respect to the 
vertex it contains. For instance, at the right-hand side of Fig.\ 
\ref{Fig1}, each straight line issued from ${\rm A}_{2}$ crosses at 
most once the stability frontier between ${\rm A}_{2}$ and the 
symmetry axis.

\noindent{\sl c)} Each instability domain is convex.
\subsection{Inverse-mass plane for unequal charges}
\label{sub:Gen-Resu:Unequal}
For arbitrary charges $q_{i}$, the threshold energy
(\ref{def-stab}) is modified
\begin{equation}
	E_{1i}^{(2)}=-{(q_{1}q_{i})^{2}\over 2(\alpha_{1}+\alpha_{i})}.
	\label{mod-stab}
\end{equation}
The separation between the two thresholds, (T), which plays a crucial 
role in the discussion, is given by
\begin{equation}
\label{sep-thresholds1}
(\alpha _{1}  + \alpha_{3}) q_{2}^{2}=
 (\alpha_{1} +\alpha_{2}) q_{3}^{2}.
\end{equation}
It is a straight line in both pictures, i.e., for fixed charges 
$q_{i}$ in the plane of inverse masses, and in the charge plane for 
fixed masses.

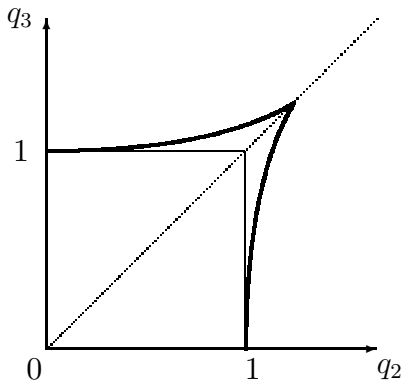
\begin{figure}[htbp] 
 \begin{center}
 \setlength{\unitlength}{0.25pt}
  \begin{picture}(1200,700)(0,0)
  \font\gnuplot=cmr10 at 12pt
  \gnuplot
  \linethickness{.7pt}
  \put(300,100){\vector(1,0){500}}
  \put(300,100){\vector(0,1){500}}
  \linethickness{.4pt}
\put(600,100){\line(0,1){300}}
\put(300,400){\line(1,0){300}}
\qbezier[120](300,100)(500,300)(800,600)
\put(270,70){\makebox(0,0)[l]{0}}
\put(600,70){\makebox(0,0)[l]{1}}
\put(250,400){\makebox(0,0)[l]{1}}
\put(800,70){\makebox(0,0)[l]{$q_{2}$}}
\put(240,600){\makebox(0,0)[l]{$q_3$}}
\linethickness{1.2pt}
 \qbezier[120](600,100)(600,350)(672,472)
 \qbezier[120](300,400)(550,400)(672,472)
  \end{picture}
  \end{center}
\caption{
\label{Fig2} 
Schematic shape of the stability domain in the Born--Oppenheimer 
limit. The heavy particles have charges $q_{2}$ and $q_{3}$. 
The charge of the light particle is set to $-q_{1}=-1$.}
\end{figure}

One expects an increase of stability near (T), where both thresholds 
become equal. This is what happens in the unit-charge case where, 
according to Hill's theorem \cite{Hill77}, we have stability for 
$\alpha_{2}=\alpha_{3}$. Another example is the Born--Oppenheimer limit with 
$m_2=m_3=\infty$, and say $m_1=q_1=1$ to fix the scales:   in the 
$(q_{2},\,q_{3})$ plane, it is observed that the stability domain 
does not extend much beyond the unit square $(q_{2}\le 1,\, q_{3}\le 
1)$, except for a spike around the $q_{2}=q_{3}$ axis, which reaches 
$q_{2}=q_{3}\simeq 1.24$ \cite{Hogreve93}. This is shown in 
Fig.~\ref{Fig2}.

For fixed charged $q_{i}$, the triangular plot of Fig.~\ref{Fig1} can be 
used again. The threshold separation (T) is a straight line passing 
through the (unphysical) point $\alpha_{1}=-1,\, 
\alpha_{2}=\alpha_{3}=1$, which is the mirror image of ${\rm A}_{1}$ with 
respect  to ${\rm A}_{2}{\rm A}_{3}$. As seen in II, each 
instability domain remains star shaped and convex, as in the case of
unit charges. In particular, if $q_{3}<q_{2}$ and $q_{3}<1$, the 
entire sub-triangle limited by (T) and ${\rm A}_{2}$ corresponds to 
stability. If furthermore $q_{2}<1$, then we have stability everywhere.

Following a suggestion by  Gribov \cite{Gribov96}, one can also 
consider level lines of constant \emph{relative} binding, i.e., such 
that
  \begin{equation}
  	E_{12}^{(2)}<E_{13}^{(2)} \quad\hbox{and}\quad 
  	E^{(3)}/E_{12}^{(2)}=\lambda>1. 
  	\label{eq:Gribov-1}
  \end{equation} 
(Remember that both $E^{(3)}$ and $E_{12}^{(2)}$ are negative.)  What 
happens is that the domain $E^{(3)}/E_{12}^{(2)}<\lambda$ (including 
the case where particle 3 is unbound, where we set 
$E^{(3)}=E_{12}^{(2)}$) is also convex and star shaped.

The proof is essentially the same as for 
the domain of instability. One first rescales the $\alpha_{i}$ from 
$\alpha_{1}+\alpha_{2}+\alpha_{3}=1$ to $\alpha_{1}+\alpha_{2}=1$, so 
that the threshold energy $E^{(2)}_{12}$ becomes constant. If two 
points $\vec{\alpha}=(\alpha_{1},\alpha_{2},\alpha_{3})$ and 
$\vec{\alpha}'=(\alpha'_{1},\alpha'_{2},\alpha'_{3})$ belong to the 
frontier of interest, that is to say
\begin{equation}
  E^{(3)}(\vec{\alpha})=E^{(3)}(\vec{\alpha}')=\lambda E_{12}^{(2)}, 
  	\label{eq:Gribov-2}
  \end{equation}
then as the $\alpha_{i}$ enter the Hamiltonian linearly, for any 
intermediate point $x \vec{\alpha}+(1-x)\vec{\alpha'}$ with $0\le 
x\le1$, one has
\begin{equation}
  E^{(3)}(x \vec{\alpha}+(1-x)\vec{\alpha}')
  \le xE^{(3)}(\vec{\alpha})+(1-x)E^{(3)}(\vec{\alpha}') =\lambda E_{12}^{(2)}. 
  	\label{eq:Gribov-3}
  \end{equation}
Similarly, a  decrease of $\alpha_{3}$ with 
$\alpha_{1}$ and $\alpha_{2}$ kept constant cannot do anything but decrease 
$E^{(3)}$ with $E^{(2)}_{12}$  unchanged: this proves the star-shape 
behaviour. 
\subsection{Convexity in the  \boldmath  $q_2^{-1},q_3^{-1}$  \unboldmath 
variables}
\label{subse:Gen-Resu:convexity} 
We return to the domain of strict instability, but now for fixed 
masses $m_{i}$ and variables charges. We fix $q_{1}=1$ and consider 
the frontier of stability in the ($1/q_{2},1/q_{3}$) plane.

First, we notice that the domain of stability is star-shaped with 
respect to the origin. Indeed, when a system of charges 
$(-1,q_{2},q_{3})$ is transformed into 
$(-1,q_{2}/\lambda^{2},q_{3}/\lambda^{2})$, with $\lambda>1$, the new system 
can be rescaled into $(-\lambda,q_{2}/\lambda,q_{3}/\lambda)$ which 
experiences the same attraction but less repulsion than the original 
system.

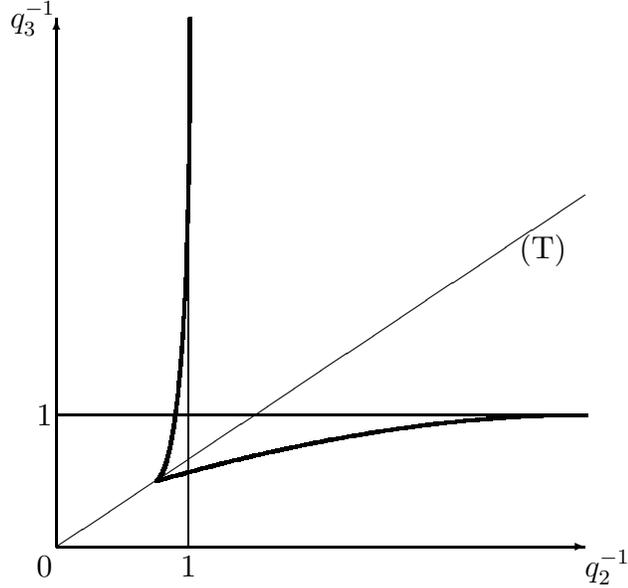
\begin{figure}[htbc]
  \vskip 1cm
 \begin{center}
 \setlength{\unitlength}{0.25pt}
  \begin{picture}(1200,900)(0,0)
  \font\gnuplot=cmr10 at 12pt
  \gnuplot
  \linethickness{.7pt}
  \put(300,100){\vector(1,0){800}}
  \put(300,100){\vector(0,1){800}}
  \put(300,100){\line(3,2){800}}
  \linethickness{.4pt}
\put(500,100){\line(0,1){800}}
\put(300,300){\line(1,0){800}}
\put(270,70){\makebox(0,0)[l]{0}}
\put(1100,70){\makebox(0,0)[l]{$q_{2}^{-1}$}}
\put(230,900){\makebox(0,0)[l]{$q_3^{-1}$}}
\put(500,70){\makebox(0,0)[c]{1}}
\put(270,300){\makebox(0,0)[l]{1}}
\put(1000,550){\makebox(0,0)[l]{(T)}}
\linethickness{1.2pt}
\qbezier(450,200)(800,300)(1100,300)
\qbezier(450,200)(500,250)(500,900)
\end{picture}
  \end{center}
\vskip .3cm 
\caption{\label{Fig3} Shape of the stability domain in 
the plane of inverse charges $(q_2^{-1},q_3^{-1})$, with normalisation 
$q_1=1$.  The lines $q_2^{-1}=1$ and $q_2^{-1}=1$ are either 
asymptotes, or part of the border, starting from a value of $q_2^{-1}$ 
or $q_3^{-1}$ which might be less than 1, unlike the case shown in 
this figure.}
\end{figure}

If $m_{2}<m_{3}$, the threshold separation (T) has a slope larger than 
unity in the ($1/q_{2},1/q_{3}$) plane.  If $m_{2}>m_{3}$, this is the 
reverse.  Consider now for definiteness the domain where %
\begin{equation}
 q_{2}^{2}\left(\alpha_{1}+\alpha_{3}\right)
 > q_{3}^{2}\left(\alpha_{1}+\alpha_{2}\right).
  	\label{eq:convex-1}
  \end{equation}
In this domain, the (1,2) atom is more bound than (1,3) and it is the 
energy of (1,2) to which $E^{(3)}$ should be compared.  The 
ground-state energy $E^{(3)}$ of the Hamiltonian (\ref{Hamiltonian}) 
is separately and globally concave in $q_{12}$, $q_{23}$ and $q_{23}$.  
With our choice of the charges we have 
\begin{equation}
\label{eq:convex-3}
q_{12}=q_{2}, \qquad q_{13}=q_{3}, \qquad q_{23}=q_{2}q_{3},
\end{equation}
but we can make a rescaling taking 
\begin{equation}
\label{eq:convex-4}
\bar{q}_{12}=1, \qquad
\bar{q}_{13}={q_{3}\over q_{2}}, \qquad
\bar{q}_{23}=q_{3}.
\end{equation}
In this way, the binding energy of (1,2) is fixed. As $E^{(3)}$ is a 
concave function of $\bar{q}_{13}$ and $\bar{q}_{23}$, the instability 
domain is convex in the  $(\bar{q}_{13},\bar{q}_{23})$ plane. This means a 
segment of  straight line 
\begin{equation}
\label{eq:convex-5}
\bar{q}_{13}=a\bar{q}_{23}+b
\end{equation}
joining two points of the domain also belongs to the domain. But this 
equation translates into
\begin{equation}
\label{eq:convex-6}
{1\over q_{2}}=a+b\left({1\over q_{3}}\right),
\end{equation}
and thus also represents a straight line in the $(1/q_{2},1/q_{3})$ 
plane. Thus each instability domain is convex in this plane.
This is schematically pictured in Fig.~\ref{Fig3}.
\section{Instability of \boldmath $(\alpha, p, \mu^-)$ \unboldmath 
systems}
\label{se:Alpha}
The problem of the instability of ions such as $(\alpha, p, \mu^-)$, 
$(\alpha, d, e^-)$, etc., has been considered by several authors.  
What has been shown is essentially that, within the Born--Oppenheimer 
approximation, the effective $(\alpha, p)$ or $(\alpha, d)$ potential 
is unable to support a bound state \cite{Chen90}.  The proof below is 
more general, for none of the masses is assumed to be very large.

First we notice that for equal charges $q_2=q_3> 1.24$ 
\cite{Hogreve93} the 3-body system is unstable for equal masses 
$\alpha_2=\alpha_3$, because it is unstable for $\alpha_2=\alpha_3=0$ 
and $\alpha_2=\alpha_3=1/2$, and the domain of instability with respect 
to a given threshold is convex in the triangle of inverse masses.

Furthermore, using for instance the $(q_2,q_3)$ plane and keeping the 
masses equal and constant, we know that the system is unstable for a 
fixed $q_2>1.24$ and $q_3$ very small, because particle 3 is almost 
free and because the system (1,2) is repulsive. So using convexity in 
$q_3$ for fixed $q_2$, we prove that the system is unstable for 
$0<q_3<q_2$, if $q_2>1.24$, for $\alpha_2=\alpha_3$.

Therefore, if $\alpha_2=\alpha_3$, a system with  $q_2=2$ and $q_3=1$ 
is unstable. But the system $(1,2)$ is repulsive and for 
$\alpha_1=\alpha_2=0$, i.e., $m_1=m_2=\infty$, we have instability. 
Now, in the left half-triangle $\alpha_2\le\alpha_3$, where
\begin{equation}
(\alpha_1+\alpha_3)q_2^2=4(\alpha_1+\alpha_3)>
(\alpha_1+\alpha_2)q_3^2=\alpha_1+\alpha_2,
\end{equation}
so that it is the $(1,2)$ system which is more negatively bound.  
Therefore, we can use the star-shape instability in the whole left 
half-triangle, which includes not only $\alpha p \mu^-$ or $\alpha p 
e^-$, but also $\alpha d \mu^-$ or $\alpha t \mu^-$. Notice that the 
proof does not work for $q_{2}>q_{3}$, $m_{3}>m_{2}$. 
\section{Some limiting configurations}
\label{se:small-q3}
\subsection{The very asymmetric Born--Oppenheimer case}
\label{subse:Born-Oppenheimer}
In II  the case  was considered where the masses $m_2$ and $m_3$ are 
both infinite, but the corresponding charges $q_2$ and $q_3$ are 
freely varying.  The stability domain is shown in Fig.~\ref{Fig2}, 
with a normalisation  $q_1=1$.  The domain is of course symmetric 
under $(q_2\leftrightarrow q_3)$ exchange, and includes the 
$(q_2\le1,q_3\le1)$ unit square.  We already mentioned the  peak at 
$q_2=q_3\simeq1.24$. The frontier starts from $q_2=1,q_3=0$, where a behaviour 
\begin{equation}
\label{end-frontier-BO}
q_2-1\simeq18{q_3\over(-\ln q_3)^3}.
\end{equation}
is  proved in II. This leading order (\ref{end-frontier-BO}) is however very crude, 
as the first corrections  differ only by terms with higher power of $(-\ln
q_3)$. 
\subsection{The Born--Oppenheimer approximation for \boldmath $m_{1}$ 
and $m_{2}$ very large, $q_{1}=q_{2}=1$ \unboldmath}
\label{subse:BO-appr}
We now consider systems analogous to ($p,\bar{p}, e^-$). In the limit 
of strictly infinite masses, we have a point source of charge 
$q_{2}-q_{1}$ acting on the charge $q_{3}>0$. Thus there is no binding 
if $q_{2}>q_{1}$ and binding for $q_{2}<q_{1}$. So, a non-trivial 
case consists of $m_{1}$ and $m_{2}$ very large but finite, and 
$q_{1}=q_{2}=1$. We argue below that binding is unlikely if 
$m_{3}q_{3}$ is very small.

With obvious notations, the adiabatic approximation  relies on the 
decomposition
\begin{eqnarray}
\label{BO-decomp}
&&H=-{\Delta\over2\mu_{12}}+h(r)-1/r\nonumber \\
&&h=-{\Delta_{3}\over2m_{3}}-{q_{3}\over r_{13}}+{q_{3}\over r_{23}}
\end{eqnarray}
with $\mu_{12}^{-1}=m_{1}^{-1}+m_{2}^{-1}$. Then $H\ge \tilde H$, 
where $\tilde H$ is deduced from $H$ by replacing $h$ by its 
ground-state energy or its infimum, say $\inf(h)$, which is a 
function of $r=r_{12}$.

The very crude inequality
\begin{equation}
\label{BO-ineg1} r_{23}\le r+r_{13} 
\end{equation}
leads to 
\begin{equation}
\label{BO-ineg2} h\ge-{\Delta_{3}\over 2 m_{3}}-{q_{3}r\over r_{13}^{2}}.
 \end{equation}
As it is known that
\begin{equation}
\label{BO-ineg3} -\Delta_{3}-{1\over 4 r_{13}^{2}}>0,
\end{equation}
we are sure that 
\begin{equation}
\label{BO-ineg4}  h\ge 0 \qquad \hbox{if}\qquad 8m_{3} q_{3} r <1.
\end{equation}
On the other hand,
\begin{equation}
\label{BO-ineg5}
h\ge -{\Delta_{3}\over 2 m_{3}}-{q_{3}\over r_{13}}\ge 
-{m_{3}q_{3}^{2}\over 2},
\end{equation}
which is obtained for $r\to\infty$. So 
\begin{equation}
\label{BO-ineg6}
\inf(h)\left\{
\begin{array}{l@{\quad}l@{\quad}l}
=0&\hbox{for}& r\le (8 m_{3}q_{3})^{-1}=R\\
\ge -m_{3}q_{3}^{2}/2&\hbox{for}& r>R.\\
\end{array}
\right.
\end{equation}
(Of course, $\inf h(r)$ must be a continuous function leaving 0 for 
some $r>R$, and reaching $ -m_{3}q_{3}^{2}/2$ at large $r$.)

Consider now the relative motion as described by
\begin{equation}
\label{BO-H-tilde}
\tilde{H}=-{\Delta\over2\mu_{12}}+\inf(h)-{1\over r}.
\end{equation}
If we neglect $\inf(h)$, $\tilde{H}$ reduces to the Schr{\"o}dinger 
equation for a two-body atom, with ground-state energy 
$-\mu_{12}/2$ and reduced radial wave function 
$u(r)=2\mu_{12}^{3/2}r\exp(-\mu_{12}r)$. For $\mu_{12}\gg 1$, this 
wave function is concentrated near $r=0$ and $\inf(h)$ can be 
considered as a perturbation. The first order correction $\delta E$ 
is negative and is such that
\begin{equation}
\label{BO-deltaE}
\vert\delta E\vert\le{m_{3}q_{3}^{2}\over 2}\int\limits_{R}^\infty 
u(r)^{2}\,{\rm d} r=\left[
{\mu_{12}^2\over 64m_{3}} + {\mu_{12}q_{3}\over 8} + 
  {m_{3}q_{3}^2\over 2}\right]\exp\left[-{\mu_{12}\over 4 m_{3}q_{3}}\right].
\end{equation}
In other words, $\delta E$ vanishes exponentially when $m_{1}$ and 
$m_{2}\to\infty$ for fixed $q_{3}$ and $m_{3}$. This is beyond the 
accuracy of the Born--Oppenheimer approximation, and strongly 
suggests that there is no binding if $m_{3}q_{3}\ll \mu_{12}$, which 
reduces, for $m_{1}=\infty$, to $m_{3}q_{3}\ll m_{2}$.
\subsection{The small \boldmath $m_{3}q_{3}$ limit for $m_1=\infty$\unboldmath}
\label{subse:small-q3-large-m1}
We now extend our study of the configurations with charges 
$q_{2}\gtrsim 1$ and $q_{3}\ll q_{2}$, with normalisation  $q_{1}=1$. 
Instead of the Born--Oppenheimer limit, we consider the  somewhat 
opposite case where $m_{1}=\infty$, i.e., the lower side ${\rm 
A}_{2}{\rm A}_{3}$  of the triangular plot.
The threshold separation (T) happens for 
\begin{equation}
\label{threshold1-large-m1}
{\alpha_3\over\alpha_2}=\left({q_3\over q_2}\right)^2,
\end{equation}
very close to A$_2$. Some crude variational calculations  has convinced us 
that the frontier occurs for
$\alpha_3/\alpha_2={\cal O}(q_3)$, not ${\cal O}(q_3^2)$.
This suggests a first order calculation in $q_3$.
We temporarily fix the scale at $\alpha_2=1$ and split the Hamiltonian into
\begin{eqnarray} 
\label{split-H-large-q1}
H&=&H_0+q_3H_1,\nonumber\\
H_0&=&-{1\over2}\Delta_{2}-{q_2\over r_2},\\
H_1&=&-{1\over2}\left({\alpha_3\over q_3}\right)
\Delta_{3}-{1\over r_3}+{q_2\over r_{23}},\nonumber
\end{eqnarray}
where $\vec{\rm r}_1=0$, and
$r_{23}=|\vec{\rm r}_3-\vec{\rm r}_2|$. We are faced with a standard problem
of degenerate perturbation theory. At zeroth order, we get the energy 
$E_{0}$ and eigenfunction $\Psi_0$
\begin{equation}
\label{zeroth-large-m1}
E_0=-{q_2^2\over2},\qquad
\Psi_0=\psi(\vec{\rm r}_2)\varphi(\vec{\rm r}_3),
\end{equation}
where $\psi(\vec{\rm r}_2)=\pi^{-1/2}(q_2)^{-3/2}
\exp-(q_2r_2)$ and
 $\varphi(\vec{\rm r}_3)$, yet unspecified, 
 is  determined by diagonalising the restriction
$\,\widetilde{\!H}_1$ of $H_1$ to the ground-state eigen\-space of $H_0$. This
reads
\begin{eqnarray}
\label{first-large-m1}
\,\widetilde{\!H}_1\varphi(\vec{\rm r}_3)&=&E_1\varphi(\vec{\rm r}_3)
\nonumber\\
\,\widetilde{\!H}_1&=&-{1\over2}\left({\alpha_3\over q_3}\right)
\Delta_{3}-{1\over r_3}+{q_2\over r_{3}}f(q_2r_3),\\
f(x)&=&1-(1+x)\exp(-2x).\nonumber
\end{eqnarray}

For $q_2<1$, the potential in (\ref{first-large-m1}) exhibits an asymptotic
Coulomb behaviour which is attractive. Thus $\,\widetilde{\!H}_1$ supports
bound states whatever inverse mass
$\alpha_3/ q_3$ is involved. We recover the property seen in (II) that for
$q_3<1$ and $q_2<1$, the 3-body system is stable for any choice of the 
constituent masses.

For $q_2\ge1$, the potential in (\ref{first-large-m1})
has a repulsive Coulomb tail or decreases exponentially. At best,
it offers a short-range pocket of attraction to trap the charge $q_3$.
The short-range character is governed by the exponential in
the form factor $f$, as per Eq.~(\ref{first-large-m1}).
Such a potential supports a bound state provided the mass
$q_3/\alpha_3$ is large enough, say $q_3/\alpha_3>\mu_c$. 
This is why at the frontier $\alpha_3={\cal O}(q_3)$.

\begin{figure}[htbp] 
 \begin{center}
 \setlength{\unitlength}{0.20pt}
  \begin{picture}(1200,850)(100,20)
  \font\gnuplot=cmr10 at 12pt
  \gnuplot
  \linethickness{.7pt}
  \put(300,100){\vector(1,0){900}}
  \put(300,100){\vector(0,1){700}}
  \linethickness{.7pt}
\qbezier(300,144)(780,408)(1100,801)
\put(270,100){\makebox(0,0)[r]{0}}
\put(270,300){\makebox(0,0)[r]{5}}
\put(270,500){\makebox(0,0)[r]{10}}
\put(270,700){\makebox(0,0)[r]{15}}
\put(300,100){\makebox(0,0){-}}
\put(300,300){\makebox(0,0){-}}
\put(300,500){\makebox(0,0){-}}
\put(300,700){\makebox(0,0){-}}
\put(300,50){\makebox(0,0){1}}
\put(700,50){\makebox(0,0){2}}
\put(1100,50){\makebox(0,0){3}}
\put(300,100){\makebox(0,0){${\scriptscriptstyle \vert}$}}
\put(500,100){\makebox(0,0){${\scriptscriptstyle \vert}$}}
\put(700,100){\makebox(0,0){${\scriptscriptstyle \vert}$}}
\put(900,100){\makebox(0,0){${\scriptscriptstyle \vert}$}}
\put(1100,100){\makebox(0,0){${\scriptscriptstyle \vert}$}}
\put(1200,50){\makebox(0,0)[r]{$q$}}
\put(270,800){\makebox(0,0)[r]{$\mu_{c}$}}
 \end{picture}
\end{center}
%
\caption{\label{Fig4} Minimal reduced mass $\mu_c$ to achieve
binding in the potential $V=-1/r+q/r(1-(1+qr)\exp(-2qr))$.}
\end{figure}
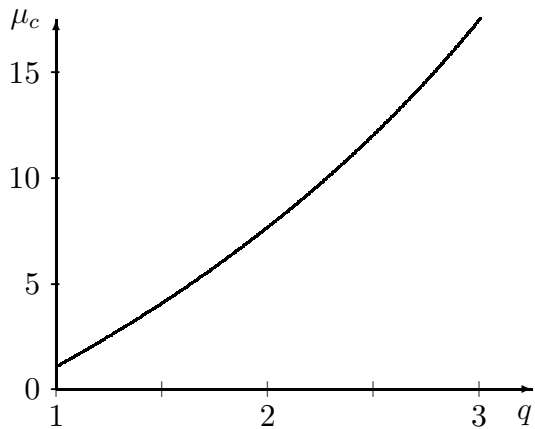

Calculating the critical mass $\mu_c$ accurately as a function of $q_2$  is a
routine numerical work. One can for instance integrate
 the radial equation at zero energy and look at whether or not a node occurs
in the radial wave-function at finite  distance. The result is shown in
Fig.~\ref{Fig4}.

The behaviour observed in Fig.\ \ref{Fig4} is not surprising.
After rescaling, the Hamiltonian of Eq.\ (\ref{first-large-m1})
can be rewritten as
\begin{equation}
\label{first-large-m1-rewritten}
\,\widetilde{\!H}_1=2q^2\left[-{1\over \mu}{{\rm d}^2\over{\rm d}r^2}-
\left({\exp(-r)\over r}+{\exp(-r)\over 2}\right)+{1-1/q\over r}\right]
\end{equation}
where $q=q_2$ and $\mu=q_{3}/\alpha_{3}$.  The critical mass for 
achieving binding in a Yukawa potential $V_{1}=-\exp(-r)/r$ is well 
known \cite{Jacksonetal} and well studied \cite{Yukawacritical}.  It 
is $\mu_{1}\simeq 1.679$.  For an exponential, it is about $1.446$ 
\cite{Jacksonetal}, and thus $\mu_{2}\simeq 2.892$ for 
$V_{2}=-\exp(-r)/2$.  It is easily seen that the critical coupling 
$\mu_{c}$ for binding in $V_{1}+V_{2}$ is such $\mu_{c}^{-1}\le 
\mu_{1}^{-1}+\mu_{2}^{-1}$.  This means $\mu_{c}\ge1.06$ for the the 
attractive part in (\ref{first-large-m1-rewritten}), a bound not very 
far from the computed value $\mu_c\simeq 1.10$.  This corresponds to 
the case  $q=1$ in Fig.\ \ref{Fig4}.  For $q>1$, the repulsive Coulomb tail 
makes it necessary to use a larger value of $\mu_c$, this explaining 
the rise observed in Fig.~\ref{Fig4}. 
\subsection{Stability frontier for small \boldmath $q_3/q_2$,
and $q_2>1$\unboldmath}
\label{subse:frontier-small-q3}

We just established that for $m_1=\infty$, small $q_3$, and
$q_2>1$, the stability frontier  lies at some
$\alpha_3\simeq q_3/\mu_c $, where $\mu_c(q_2)$ 
is computable from a simple radial equation. We now study
how the frontier behaves as $m_1$ becomes finite. We are
near A$_2$ in the triangle, where $\alpha_2\simeq1$, and
$\alpha_1$ and $\alpha_3$ are small.

We introduce the Jacobi variables
\begin{equation}
\label{Jacobi}
\vec\rho=\vec{\rm r}_2-\vec{\rm r}_1,\quad
\vec\lambda=\vec{\rm r}_3-{\alpha_2\vec{\rm r}_1
+\alpha_1\vec{\rm r}_2\over\alpha_1+\alpha_2},
\end{equation}
in terms of which the relative distances are
\begin{equation}
\label{relative-distances}
\vec{\rm r}_{12}=\vec\rho,\quad
\vec{\rm r}_{23}=\vec\lambda-
{\alpha_2\over\alpha_1+\alpha_2}\vec\rho,\quad
\vec{\rm r}_{31}=-\vec\lambda-
{\alpha_1\over\alpha_1+\alpha_2}\vec\rho,
\end{equation}
and the Hamiltonian reads
\begin{eqnarray}
\label{relative-H-1}
H=&&-{1\over2}(\alpha_1+\alpha_2)\Delta_\rho
-{q_2\over\rho}-{1\over2}\left(\alpha_3+
   {\alpha_1\alpha_2\over\alpha_1+\alpha_2}\right)\Delta_\lambda
\nonumber\\
&&-{q_3\over\vert\vec\lambda+\alpha_1\vec\rho
       /(\alpha_1+\alpha_2)\vert}
+{q_2q_3\over\vert\vec\lambda-\alpha_2\vec\rho
       /(\alpha_1+\alpha_2)\vert},
\end{eqnarray}
besides the centre-of-mass motion, which will be now omitted.

A first rescaling 
$\vec\rho\rightarrow(\alpha_1+\alpha_2)\vec\rho/\alpha_2$
results into
\begin{eqnarray}
\label{relative-H-2}
H=&&-{1\over2}\left({\alpha_2^2\over\alpha_1+
\alpha_2}\right)\Delta_\rho 
-{\alpha_2\over\alpha_1+\alpha_2}{q_2\over\rho}\nonumber\\
&&-{1\over2}\left(\alpha_3+
   {\alpha_1\alpha_2\over\alpha_1+\alpha_2}\right)\Delta_\lambda
-{q_3\over\vert\vec\lambda+\alpha_1\vec\rho
       /\alpha_2\vert}
+{q_2q_3\over\vert\vec\lambda-\vec\rho\vert},
\end{eqnarray}
which is the scale transformed of
\begin{equation}
\label{relative-H-3}
\,\overline{\!H}=-{1\over2}\bar\alpha_2\Delta_{2}
-{1\over2}\bar\alpha_3\Delta_{3}-{\bar q_2\over r_2}
+{\bar q_2\bar q_3\over r_{23}}-{\bar q_3\over
\vert\vec{\rm r}_3+\alpha_1\vec{\rm r}_2/\alpha_2\vert},
\end{equation}
provided the inverse masses in $\,\overline{\!H}$ are proportional to these
in $H$, and the strengths in $\,\overline{\!H}$ to these of $H$.
A convenient rule of transformation of masses is
\begin{eqnarray}
\label{alpha-bar}
\bar \alpha_2&=&{1\over\alpha_2+\alpha_3}
\left({\alpha_2^2\over\alpha_1+\alpha_2}\right)\nonumber\\
\bar \alpha_3&=&{1\over\alpha_2+\alpha_3}
\left(\alpha_3+{\alpha_1\alpha_2\over\alpha_1+\alpha_2}
\right),
\end{eqnarray}
since it changes our triangular normalisation $\sum\alpha_i=1$
into $\bar\alpha_2+\bar\alpha_3=1$.

\begin{figure}[htbc]
  \begin{center}
 \setlength{\unitlength}{0.25pt}
  \begin{picture}(1200,690)(0,20)
  \font\gnuplot=cmr10 at 12pt
  \gnuplot
  \put(300,100){\line(1,0){600}}
  \put(300,100){\line(3,5){300}}
  \put(900,100){\line(-3,5){300}}
  \put(590,630){\makebox(0,0)[l]{A$_1$}}
  \put(910,70){\makebox(0,0)[l]{A$_2$}}
  \put(270,70){\makebox(0,0)[l]{A$_3$}}
 \put(750,100){\line(3,5){75}}
\put(850,100){\line(2,5){20}}
  \put(720,180){\makebox(0,0)[l]{(F)}}
  \put(790,140){\makebox(0,0)[l]{(T)}}
 \end{picture}
 \end{center}
\caption{
\label{Fig5} 
Expected behaviour of the frontier (F) in a situation where $q_2>1$ 
and the ratio $q_3/q_2$ is very small, so that the threshold 
separation (T) is very close to A$_2$.}
\end{figure}
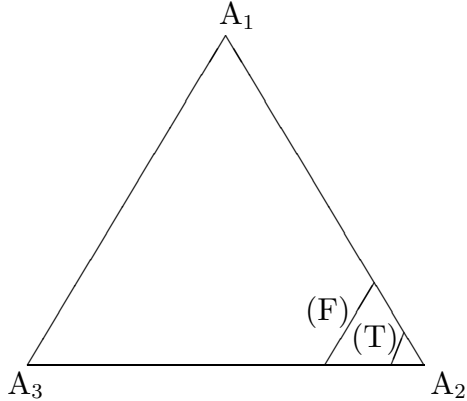

For the charges, the simultaneous identification
\begin{eqnarray}
\label{q-bar-1}
\bar q_2&=&Cq_2{\alpha_2\over\alpha_1+\alpha_2},\nonumber\\
 \bar q_3&=&Cq_3,\\
\bar q_2\bar q_3&=&Cq_3q_2,\nonumber
\end{eqnarray}
results into
\begin{equation}
\label{q-bar-2}
\bar q_2=q_2,\qquad \bar q_3=q_3{\alpha_1+\alpha_2\over \alpha_{2}}.
\end{equation}
The rescaled Hamiltonian (\ref{relative-H-3}) slightly differs from the
Hamiltonian (\ref{split-H-large-q1}) corresponding to $m_1=\infty$.
 However, the difference between 
$\vert\vec{\rm r}_3+\alpha_1\vec{\rm r}_2/\alpha_2\vert$
and $r_3$ is of first order in $q_3$, and thus enters at order
$q_3^2$ in $\,\overline{\!H}$. We are then allowed to write the
frontier condition as in the previous section, namely
\begin{equation}
\label{frontier-1}
\bar\alpha_3\simeq {\bar q_3\over\mu_c},
\end{equation}
which, when translated into the original variables, reads, at
first order
\begin{equation}
\label{frontier-2}
\alpha_1+\alpha_3\simeq{ q_3\over\mu_c},
\end{equation}
to be compared with the threshold separation
$\alpha_1+\alpha_3=q_3^2$. This means the frontier is 
at first approximation a straight line,
parallel to the side A$_3$A$_1$ of the triangle of inverse masses, as schematically 
pictured in Fig.\ \ref{Fig5}.
Note that the actual frontier is certainly curved, since the instability
domains are convex, as reminded in Sec.~\ref{se:Gen-resu}.
\subsection{The small \boldmath $q_3$ limit in the $(q_2,q_3)$
plane\unboldmath}
\label{subse:small-q3-in-q2-q3-plane}
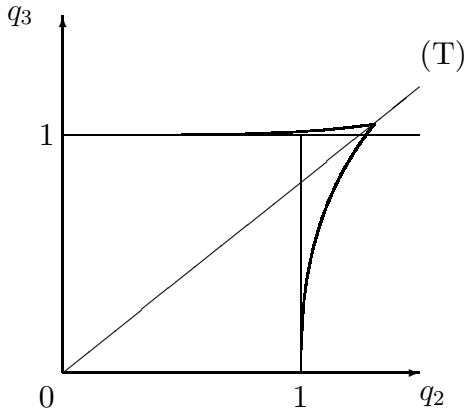
\begin{figure}[htcb]
 \begin{center}
 \setlength{\unitlength}{0.30pt}
  \begin{picture}(700,600)(100,20)
  \font\gnuplot=cmr10 at 12pt
  \gnuplot
  \linethickness{.7pt}
  \put(300,100){\vector(1,0){450}}
  \put(300,100){\vector(0,1){450}}
  \linethickness{.4pt}
\put(600,100){\line(0,1){300}}
\put(300,400){\line(1,0){450}}
\put(300,100){\line(5,4){450}}
\put(270,70){\makebox(0,0)[l]{0}}
\put(750,70){\makebox(0,0)[l]{$q_{2}$}}
\put(230,550){\makebox(0,0)[l]{$q_3$}}
\put(600,70){\makebox(0,0)[c]{1}}
\put(270,400){\makebox(0,0)[l]{1}}
\put(750,500){\makebox(0,0)[l]{(T)}}
\linethickness{.7pt}
 \qbezier(600,130)(600,300)(691.875,413.5)
 \qbezier(450,400)(600,400)(691.875,413.5)
  \end{picture}
  \end{center}
\caption{
\label{Fig6} 
Shape of the stability domain in the $(q_2,q_3)$ plane.  The frontier 
of stability leaves the vertical side $q_{2}=1$ of the unit square at 
some finite value of $q_3$.  It also leaves the horizontal line 
$q_{3}=1$ at some finite value of $q_{2}$, which can either smaller 
(as in this figure) or larger than 1.}
\end{figure}

Let us consider now the $(q_2,q_3)$ plane (with $q_1=1$) for fixed masses.
The shape of the stability domain is shown in Fig.\ \ref{Fig6}.
The frontier of stability leaves the unit square at some finite
value of $q_3$. Consider, indeed, $q_2=1$, with $\alpha_1=0$ for simplicity,
and a mass scale fixed at $\alpha_2=1$. In a (variational) approximation of
a (1,2) atom times a function describing the motion of the third particle,
we can read the calculation of  Sec.~\ref{subse:small-q3-large-m1} as
\begin{equation}
\label{q_3-mini-1}
1>q_3>\mu_c(1)\alpha_3\simeq 1.10\alpha_3
\end{equation}
being a sufficient condition for stability.

A necessary condition of the same type, i.e., $q_3>\tilde{\mu}_c\alpha_3$
can be obtained using the method of Glaser et al.\ \cite{Glaser76}.
The decomposition
\begin{equation}
\label{decomp-proj}
H=\left(H-{q_{2}q_{3}\over r_{23}}\right)+{q_{2}q_{3}\over r_{23}}
\end{equation}
yields the operator inequality \cite{Thirring79}
\begin{equation}
\label{ineg-proj}
H\ge H'=\left(H-{q_{2}q_{3}\over 
r_{23}}\right)+q_{2}q_{3}P(P\,r_{23}\,P)^{-1}P,
\end{equation}
where $P$ is the projector over the ground-state of 
$H_{0}=-\Delta_{2}-q_{2}/r_{2}$ (times the identity in the 
variable $\vec{\rm r}_{3}$). Now $H'$ is the sum of $H_{0}$ in 
the variable $\vec{\rm r}_{2}$, and 
\begin{equation}
\label{pot-in-r3-proj}
H'_{0}=-\alpha_{3}\Delta_{3}-q_{3}/r_{3} + {q_{2}q_{3}\over r_{3}}
\left[1-{x^{2}\over 1-(1+x/2)\exp(-2x)}\right]^{-1},
\end{equation}
where $x=q_{2}r_{3}$, in the variable $\vec{\rm r}_{3}$. For $q_{2}=1$,
this  potential supports a bound-state provided 
$q_3>\tilde{\mu}_c\alpha_3$, with $\tilde{\mu}_{c}>0.34$ from the  
Jost--Pais--Bargmann rule \cite{Jost}, and $\tilde{\mu}_{c}\simeq 0.64$ from a 
numerical calculation (looking for nodes in the radial wave function 
at zero energy).

If $\alpha_1>0$, a reasoning similar to that of
Subsec.~\ref{subse:frontier-small-q3} shows that the sufficient 
condition (\ref{q_3-mini-1}) is replaced by 
\begin{equation}
\label{q_3-mini-2}
q_3>\mu_c(1)(\alpha_1+\alpha_3)\simeq 1.10(\alpha_1+\alpha_3),
\end{equation}
where the normalisation is $\alpha_1+\alpha_2+\alpha_3=1$.

The result (\ref{q_3-mini-2}) is of course expected to be better if the
computed $q_3$ is small, i.e., if $\alpha_1+\alpha_3\ll 1$.

\subsection{Frontier in the \boldmath$(q_2,q_3)$ plane at small
$\alpha_3/\alpha_2$\unboldmath}
\label{q2q3map}
We remain in  the $(q_2,q_3)$ plane for fixed masses.
We assume $\alpha_1=0$ for simplicity, but some results do not depend
on this assumption. We can normalise to $\alpha_2=1$. In the limit
where $\alpha_3$ is small, the threshold separation (T), 
as given by Eq.\ (\ref{threshold1-large-m1}),
has a very small slope with respect to the $q_3$ axis.
 
The frontier exits out of the unit square at $q_2=1$ and a finite value of
$q_3$ which is close to $\alpha_3\mu_c(1)$, where $\mu_c(1)\simeq 1.10$, 
according to our previous computation. If we look at the frontier outside the 
unit square, we have two questions:
\begin{itemize}
\item[i)] in the lower part of the plot, is the frontier strictly below (T) ?
\item[ii)] in the upper part, does the frontier overcome the
line $q_3=1$ ?
\end{itemize}

\subsubsection{Lower part of the frontier}
\label{Lower-part}
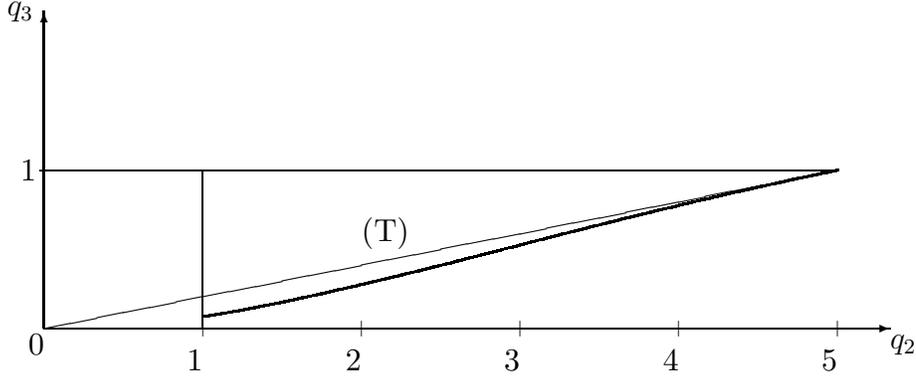
\begin{figure}[htcb]
 \begin{center}
 \setlength{\unitlength}{0.20pt}
  \begin{picture}(1900,800)(200,20)
  \font\gnuplot=cmr10 at 12pt
  \gnuplot
  \linethickness{.7pt}
  \put(300,100){\vector(1,0){1600}}
  \put(300,100){\vector(0,1){600}}
  \linethickness{.3pt}
\put(600,100){\line(0,1){300}}
\put(300,400){\line(1,0){1500}}
\put(300,100){\line(5,1){1500}}
\put(270,70){\makebox(0,0)[l]{0}}
\put(1900,70){\makebox(0,0)[l]{$q_{2}$}}
\put(230,700){\makebox(0,0)[l]{$q_3$}}
\put(600,40){\makebox(0,0)[r]{1}}
\put(600,100){\makebox(0,0){${\scriptscriptstyle\vert}$}}
\put(900,40){\makebox(0,0)[r]{2}}
\put(900,100){\makebox(0,0){${\scriptscriptstyle\vert}$}}
\put(1200,40){\makebox(0,0)[r]{3}}
\put(1200,100){\makebox(0,0){${\scriptscriptstyle\vert}$}}
\put(1500,40){\makebox(0,0)[r]{4}}
\put(1500,100){\makebox(0,0){${\scriptscriptstyle\vert}$}}
\put(1800,40){\makebox(0,0)[r]{5}}
\put(1800,100){\makebox(0,0){${\scriptscriptstyle\vert}$}}
\put(270,400){\makebox(0,0){1}}
\put(300,400){\makebox(0,0){-}}
\put(900,280){\makebox(0,0)[l]{(T)}}
\linethickness{.7pt}
\qbezier(600,123)(780,150)(1166,250)
\qbezier(1800,400)(1554,351)(1166,250)
  \end{picture}
  \end{center}
\caption{\label{Fig7} Upper bound (\protect\ref{approx-frontier}) on 
the lower part of the stability frontier, touching the threshold 
separation (T) for $q_3=1$.  A value $\alpha_3/\alpha_2=1/25$ is 
assumed here.  Note that this bound is not expected to be a good 
approximation for small $q_2$, as it does not delimit a convex domain 
of instability.}
\end{figure}

To answer the first question, let us consider a situation where $q_3$ 
is close to but smaller than 1, and thus $q_2\sim\alpha_3^{-1/2}$ is 
large.  If particles 2 and 3 would ignore each other, they would bind 
around particle 1 with approximately the same energy, since we are 
close to (T), but with different Bohr radii $R_i$, namely 
$R_3/R_2\sim\alpha_3^{1/2}\ll1$.  This suggests the approximation of a 
localised (1,3) source attracting the charge $q_{2}$, corresponding to 
a 3-body energy 
\begin{equation}
\label{point-approximation}
E_3=-{q_3^2\over\alpha_3}-q_2^2(1-q_3)^2,
\end{equation}
whose equality with the threshold $E_2=-q_2^2$ 
gives the approximate frontier
\begin{equation}
\label{approx-frontier}
q_2^2={1\over \alpha_3}{q_3\over2-q_3}
\end{equation}
which just touches (T) at $q_3=1$, as seen in Fig.\ \ref{Fig7}.
Now this approximation corresponds to write a decomposition
\begin{eqnarray}
\label{decomp-H}
H&=&H_{23}+V_{23}\nonumber\\
&=&\left[{\alpha_3\over2}\vec{\rm p}_3^2-{q_3\over r_3}
+{1\over2}\vec{\rm p}_2^2-{q_2(1-q_3)\over r_2}\right]
+q_2q_3\left({1\over r_{23}}-{1\over r_2}\right),
\end{eqnarray}
and neglect the second  term, $V_{23}$. The spherically-symmetric ground-state
$\Psi_0$ of $H_{23}$ can be chosen as a trial
variational wave-function for $H$. The Gauss theorem implies that
$\langle\Psi_0|V_{23}|\Psi_0\rangle<0$. Hence the
ground state of $H$ lies below that of
$H_{23}$, and the actual frontier is below the approximation
(\ref{approx-frontier}), therefore below (T) as long as $q_3<1$.

\subsubsection{Upper part}
\label{Upper-part}
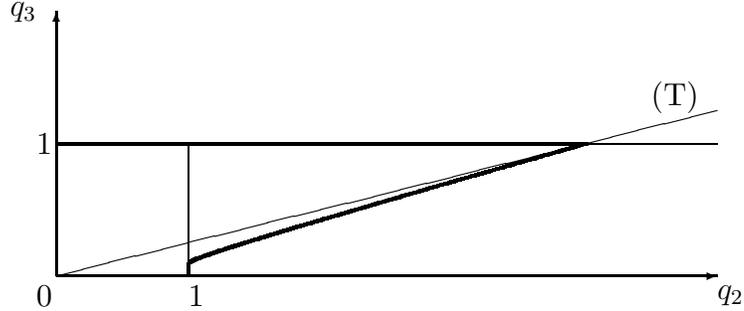
\begin{figure}
 \begin{center}
 \setlength{\unitlength}{0.25pt}
  \begin{picture}(1300,600)(0,0)
  \font\gnuplot=cmr10 at 12pt
  \gnuplot
 \linethickness{.7pt}
  \put(300,100){\vector(1,0){1000}}
  \put(300,100){\vector(0,1){400}}
  \linethickness{.4pt}
\put(500,100){\line(0,1){200}}
\put(300,300){\line(1,0){1000}}
\put(300,100){\line(4,1){1000}}
\put(270,70){\makebox(0,0)[l]{0}}
\put(1300,70){\makebox(0,0)[l]{$q_{2}$}}
\put(230,500){\makebox(0,0)[l]{$q_3$}}
\put(500,70){\makebox(0,0)[l]{1}}
\put(270,300){\makebox(0,0)[l]{1}}
\put(1200,370){\makebox(0,0)[l]{(T)}}
\linethickness{1pt}
 \qbezier(500,120)(500,130)(1100,300)
\put(500,100){\line(0,1){20}}
\put(300,300){\line(1,0){800}}
  \end{picture}
  \end{center}
\caption{
\label{Fig8} 
Schematic shape  of the stability domain in the 
$(q_2,q_3)$ plane, for the limiting case where $\alpha_3/\alpha_2$ is 
very small. Our constraints cannot exclude a very tiny peak along (T) 
above $q_{3}=1$.} 
\end{figure}

  We now turn to the question of possible binding above the line 
  $q_{3}=1$.  We restrict ourselves to $m_{1}=\infty$, although we 
  suspect that our results are more general.  Numerical investigations 
  using the method described in Appendix~B suggest the following 
  pattern.  For $m_{2}=m_{3}$, a spike is observed on the diagonal.  
  It reaches about $q_{2}=q_{3}=1.24$ in Fig.~\ref{Fig2}, 
  corresponding to $m_{2}=m_{3}=\infty$, and about $q_{2}=q_{3}=1.098$ 
  \cite{Baker90} for $m_{2}=m_{3}\ll m_{1}$.  The spike remains for 
  moderate values of the mass ratio $m_{3}/m_{2}$, as schematically 
  shown in Fig.~\ref{Fig6}.  When, however, $m_{3}/m_{2}$ exceeds a 
  value which is about 1.8, no spike is seen within the accuracy of 
  our calculations, i.e., the frontier seemingly coincides with the 
  line $q_{3}=1$, until it reaches (T), as pictured in 
  Fig.~\ref{Fig8}. 
  
We are able to show rigorously below that, for large values of 
$m_{3}/m_{2}$ no binding occurs above $q_{3}=1$ for 
$q_{2}\le (3/4)^{1/2}(m_{3}/m_{2})^{1/2}$.  Nothing can be said however 
from this latter value to $q_{2}=(m_{3}/m_{2})^{1/2}$ on the threshold 
separation (T).  In other words, a very tiny peak along (T) overcoming 
$q_{3}=1$ cannot be excluded.

For $m_{1}=\infty$ and $q_{3}=1$, the Hamiltonian reduces to 
\begin{eqnarray}
\label{abs-H}
H&=&\left[{ \vec{\rm p}_{3}^{2}\over 2m_{3}}-{1\over r_{3}}\right]+
   { \vec{\rm p}_{2}^{2}\over 2 m_{2}}-
           {q_{2}\over r_{2}} +{q_{2}\over r_{23}}\nonumber\\ 
&=&H_{0}+ 
   { \vec{\rm p}_{2}^{2}\over 2 m_{2}} -
            {q_{2}\over r_{2}} +{q_{2}\over r_{23}}.  
\end{eqnarray} 
Let $P$ be the projector on $\Phi(r_{3})$, the ground state of 
$H_{0}={ \vec{\rm p}_{3}^{2}/( 2 m_{3})}
-{1/ r_{3}}$, and $\Psi$ the ground state of $H$. We have the 
inequality 
\begin{equation}
\label{abs-eq1}
\left(\int{\rm d}\vec{\rm r}_{3}\Psi{1\over r_{23}}\Psi\right)
\left(\int{\rm d}\vec{\rm r}_{3}\Phi\,r_{23}\,\Phi\right)\ge
\left(\int{\rm d}\vec{\rm 
r}_{3}\Psi\Phi\right)^{2}=\left(P\Psi(\vec{\rm r}_{2})\right)^{2}.
\end{equation}
To estimate $\langle\Psi\vert H\vert\Psi\rangle$, we first need
 $\langle\Psi\vert H_{0}\vert\Psi\rangle$, where each $\Psi$ can be 
 read as $P\Psi+(1-P)\Psi$. We have
\begin{eqnarray}
\label{abs-eq2}
&\langle \Psi P\vert H_{0}\vert (1-P)\Psi\rangle=0,\nonumber\\
&\langle \Psi P\vert H_{0}\vert P\Psi\rangle=
\langle \Psi P\vert P\Psi\rangle\, (-{m_{3}/2}),\\
&\langle \Psi(1-P)\vert H_{0}\vert(1-P)\Psi\rangle\ge
\langle \Psi (1-P)\vert(1- P)\Psi\rangle\, (-{m_{3}/8}),\nonumber
\end{eqnarray}
and similarly, for $h_{0}={ \vec{\rm p}_{2}^{2}/( 2 m_{2})}
-{q_{2}/ r_{2}}$,
\begin{eqnarray}
\label{abs-eq3}
&&\langle \Psi P\vert h_{0}\vert (1-P)\Psi\rangle=0,\\
&&\langle \Psi(1-P)\vert h_{0}\vert(1-P)\Psi\rangle\ge
\langle \Psi (1-P)\vert(1- P)\Psi\rangle\, 
(-q_{2}^{2}{m_{2}/2}).\nonumber
\end{eqnarray}
Thus
\begin{eqnarray}
\label{abs-eq4}
\langle\Psi\vert H\vert\Psi\rangle \ge &&\vert\!\vert P\Psi\vert\!\vert^{2} 
\left(-{m_{3}\over 2}\right) +
     \langle \Psi P\vert \tilde{h}_{0}\vert P\Psi\rangle\\
&&{}+\vert\!\vert(1-P)\Psi\vert\!\vert^{2}\left[-{m_{3}\over 
8}-{q_{2}^{2}m_{2}\over 2}\right],
\end{eqnarray}
where
\begin{equation}
\label{abs-eq5}
\tilde{h}_{0}={ \vec{\rm p}_{2}^{2}/( 2 m_{2})}
-{q_{2}/ r_{2}}+ {q_{2}\over \int\Phi^{2}\, r_{23}\, {\rm d}\vec{\rm 
r}_{2}}.
\end{equation}
In a situation where $\tilde{h}_{0}$ does not support any bound state,
\begin{equation}
\label{abs-eq6a}
\langle\Psi\vert H\vert\Psi\rangle > 
\vert\!\vert P\Psi\vert\!\vert^{2} \left(-{m_{3}\over 2}\right) + 
\vert\!\vert(1-P)\Psi\vert\!\vert^{2}\left[-{m_{3}\over 
8}-{q_{2}^{2}m_{2}\over 2}\right],
\end{equation}
and hence
\begin{equation}
\label{abs-eq6b}
\langle\Psi\vert H\vert\Psi\rangle > \inf\left\{
\begin{array}{l}
-m_{3}/2\\
-m_{3}/8-m_{2}q_{2}^{2}/2.
\end{array}
\right.
\end{equation}
Now the hamiltonian $\tilde{h}_{0}$ has been studied in 
Ref.~\cite{Glaser76}, and shown not to bind if 
\begin{equation}
\label{abs-eq7}
{2m_{2}q_{2}\over m_{3}}< 1.2706.
\end{equation}
Therefore if
\begin{equation}
\label{abs-eq8}
-{m_{3}\over 2}<
-{m_{3}\over 8}-{m_{2}q_{2}^{2}\over 2}, \qquad\hbox{and}\qquad
{2m_{2}q_{2}\over m_{3}}< 1.2706,
\end{equation}
the system is unstable (the first condition implies 
$-m_{3}/ 2<-m_{2}q_{2}^{2}/2$, i.e., (1,3) is the lowest threshold).
For large $m_{3}/m_{2}$, the first condition is more constraining, so 
we have no stability above $q_{3}=1$ from  $q_{2}=0$ to $q_{2}=(3 m_{3}/(4 
m_{2}))^{1/2}$. 

\subsection{Numerical results}
\label{Subse:numeric}
We now display an estimate of the domain of stability in the 
$(q_{2},\,q_{3})$ plane, with normalisation $q_{1}=1$. The method, 
described in Appendix B, is variational. Therefore, the approximate 
domain drawn here is included in the true domain.

Our investigations correspond to $m_{1}=\infty$, and the mass ratio 
$m_{3}/m_{2}$ having the values 1, 1.1, 1.5 and 2. In each 
case, we show the whole domain, and an enlargement of its most 
interesting part, the spike above $q_{2}=1$ and $q_{3}=1$.

\begin{figure}[htbc]
\setbox1=\vbox{\includegraphics*[width=8cm]{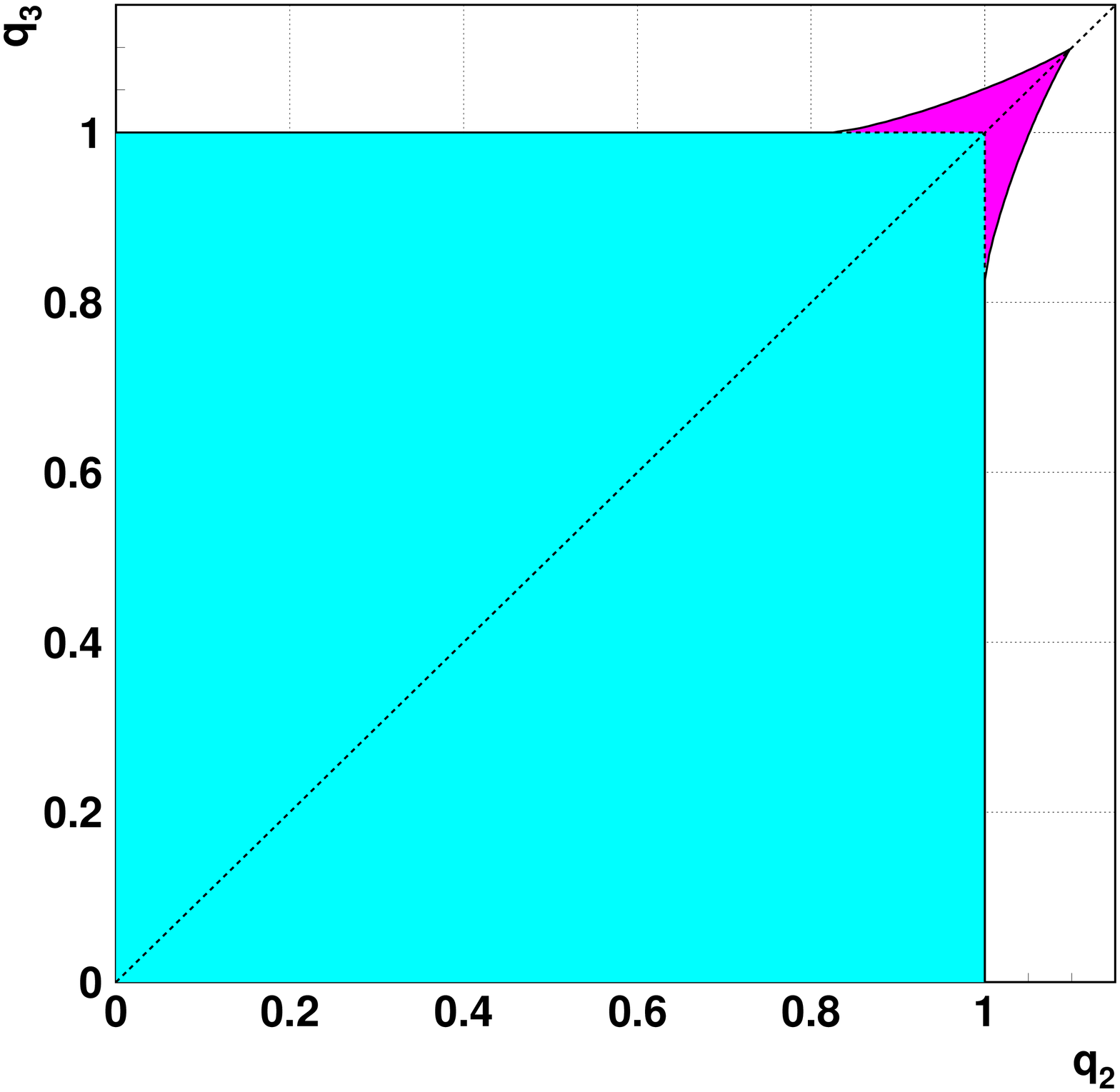}}
\setbox2=\vbox{\includegraphics*[width=8cm]{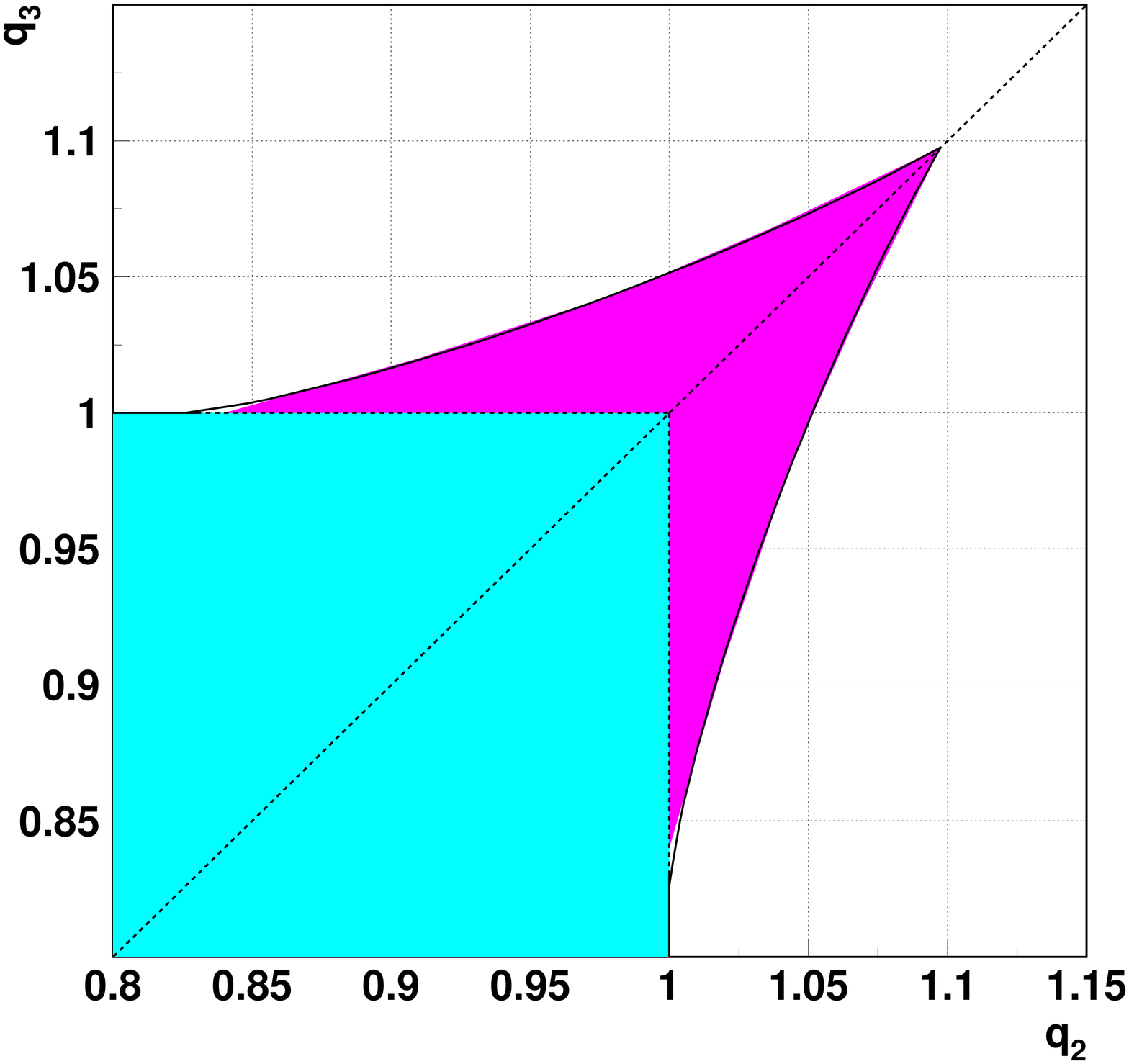}}
\centerline{\hglue 5.5cm\box1\hglue -4.5cm\box2}    
\caption{Variational estimate of the domain of stability for 
$m_{1}=\infty$ and $m_{2}/m_{3}=1$, full view (left) and enlargement 
of the spike (right).  The dotted line is the threshold separation 
(T).}
\label{Fig9}
\end{figure}
\begin{figure}[htbc]
\setbox1=\vbox{\includegraphics*[width=8cm]{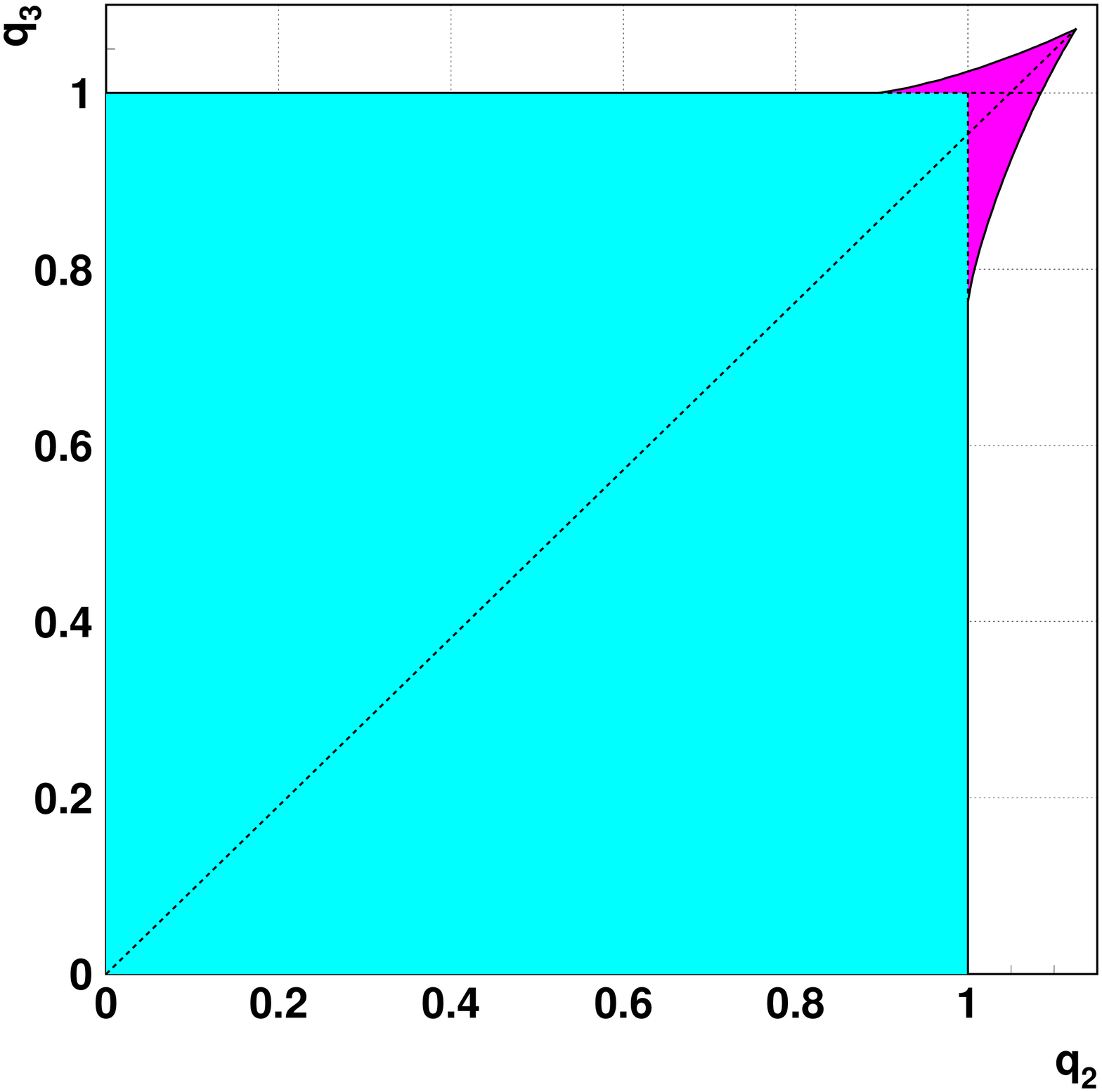}}
\setbox2=\vbox{\includegraphics*[width=8cm]{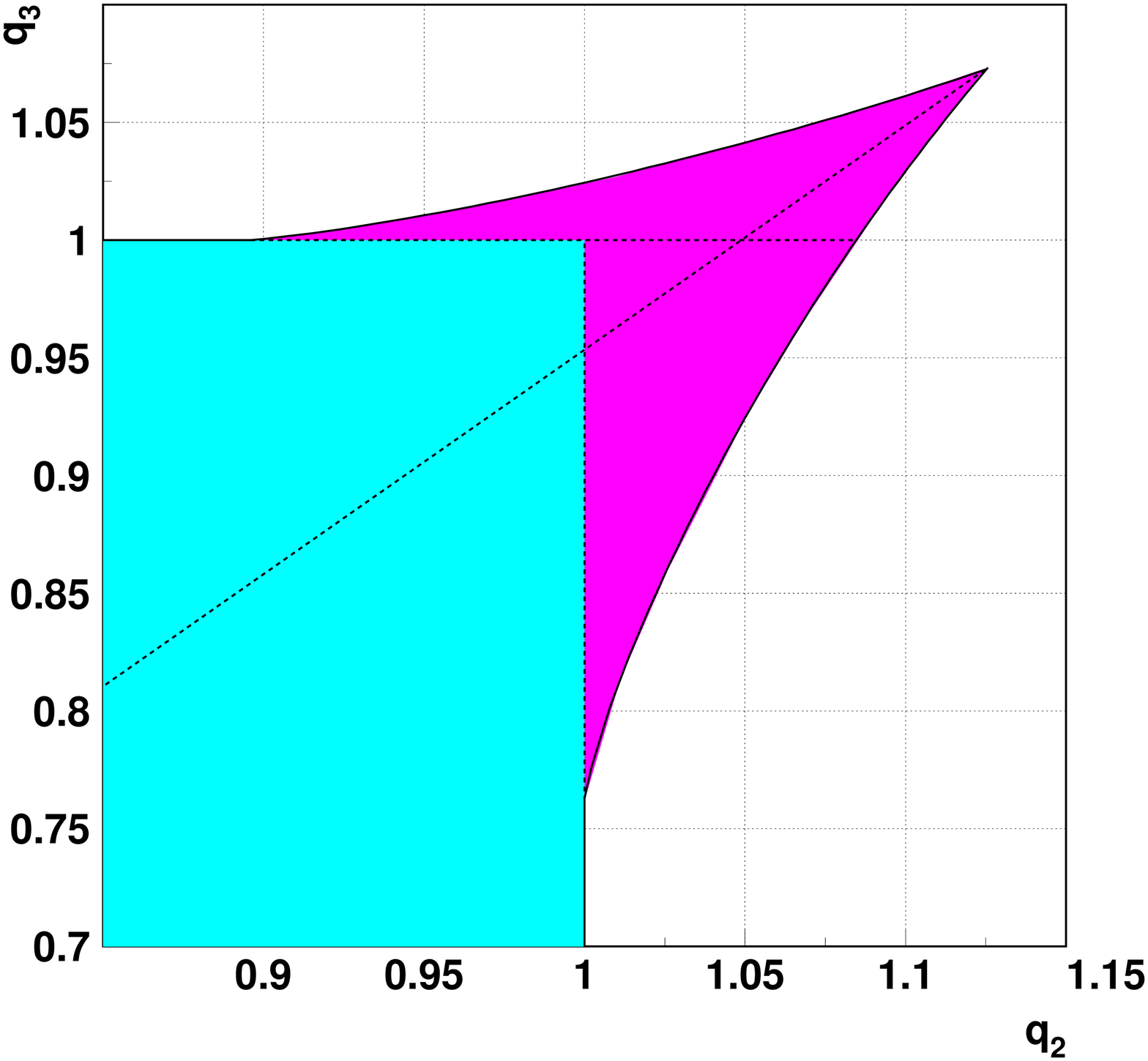}}
\centerline{\hglue 5.5cm\box1\hglue -4.5cm\box2}    
\caption{Same as Fig.~\protect\ref{Fig9}, for $m_{2}/m_{3}=1.1$}           
\label{Fig10}
\end{figure}
\begin{figure}[htbc]
\setbox1=\vbox{\includegraphics*[width=8cm]{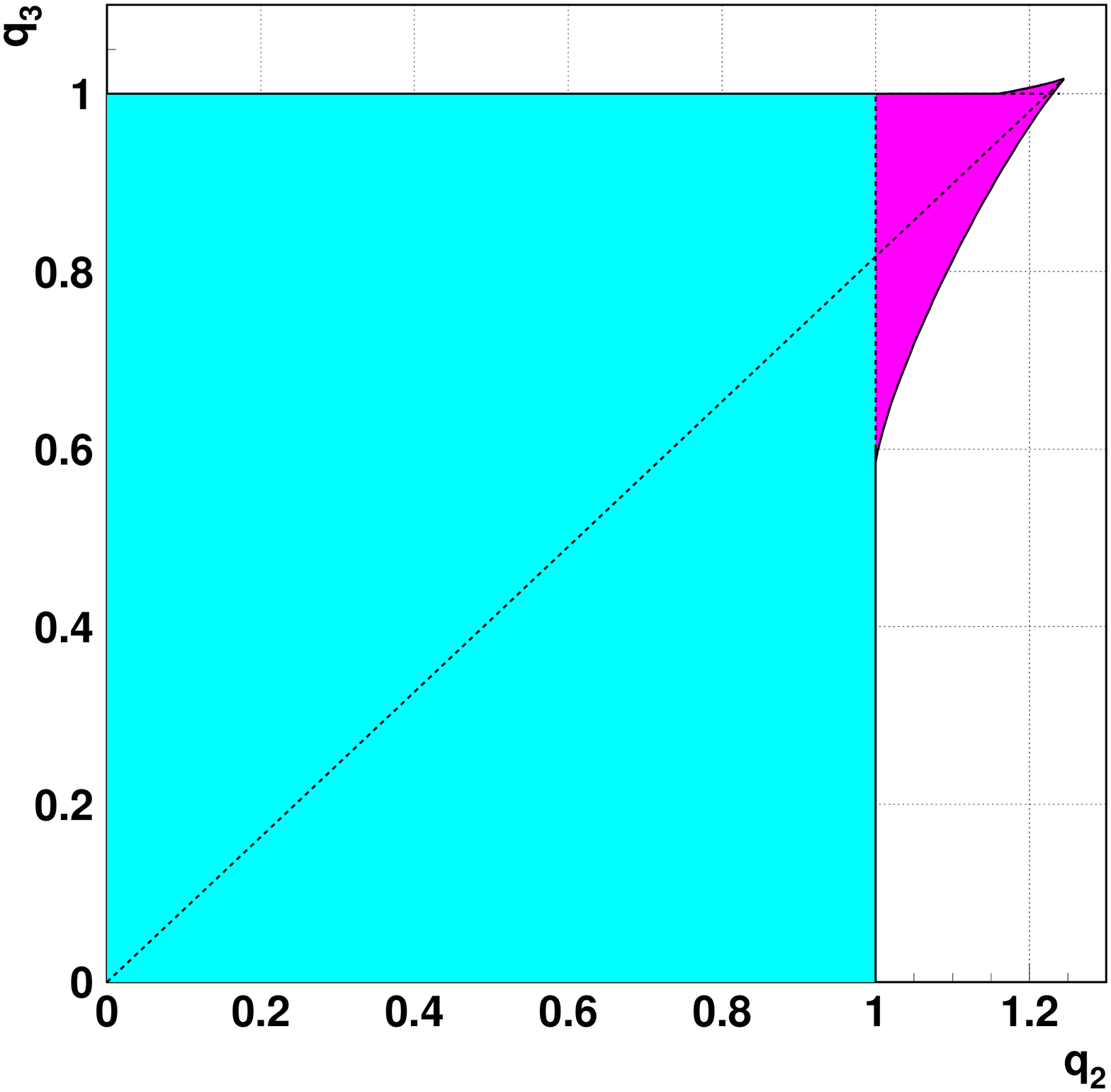}}
\setbox2=\vbox{\includegraphics*[width=8cm]{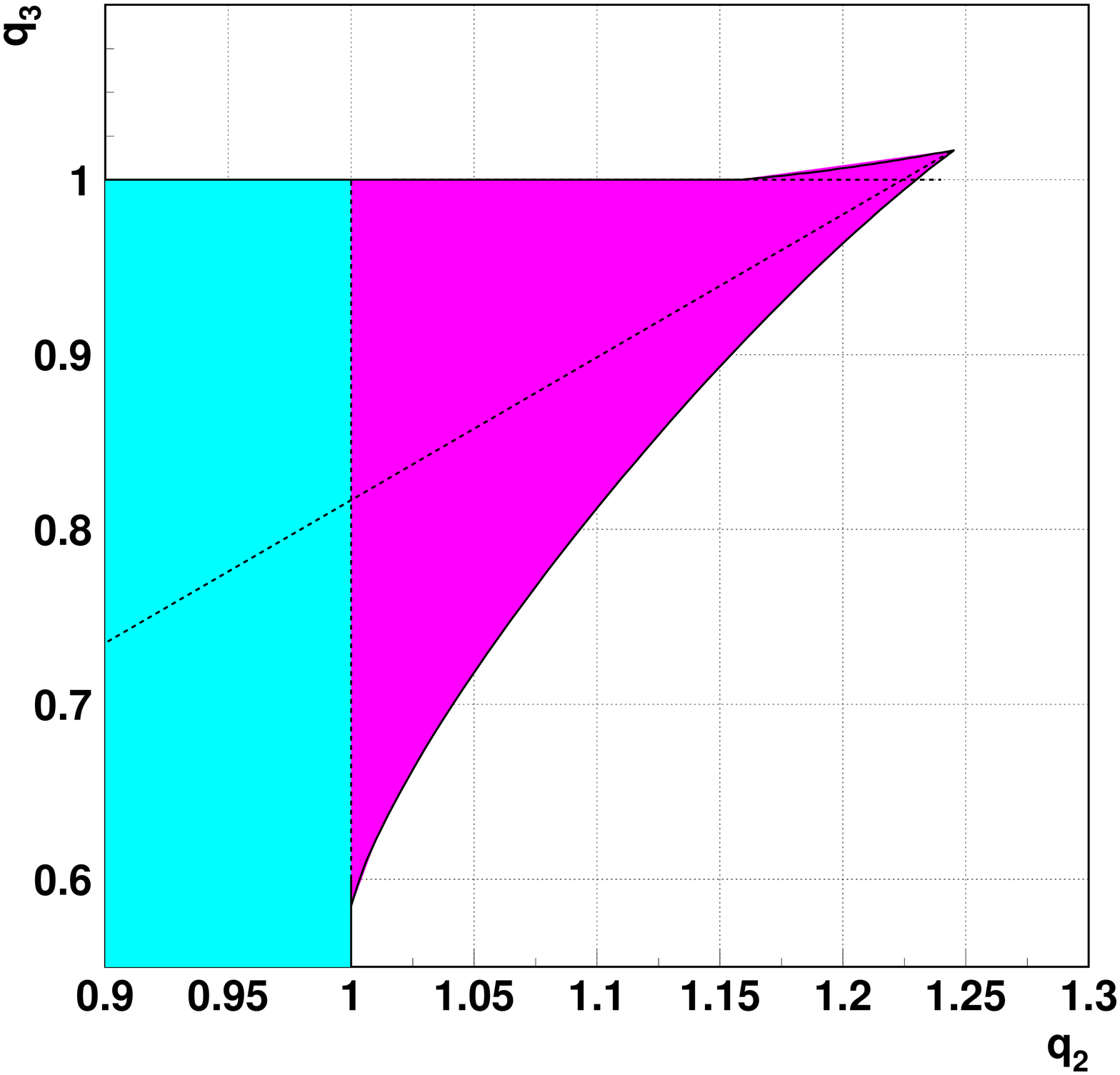}}
\centerline{\hglue 5.5cm\box1\hglue -4.5cm\box2}    
\caption{Same as Fig.~\protect\ref{Fig9}, for $m_{2}/m_{3}=1.5$}           
\label{Fig11}
\end{figure}
\begin{figure}[h]
\setbox1=\vbox{\includegraphics*[width=8cm]{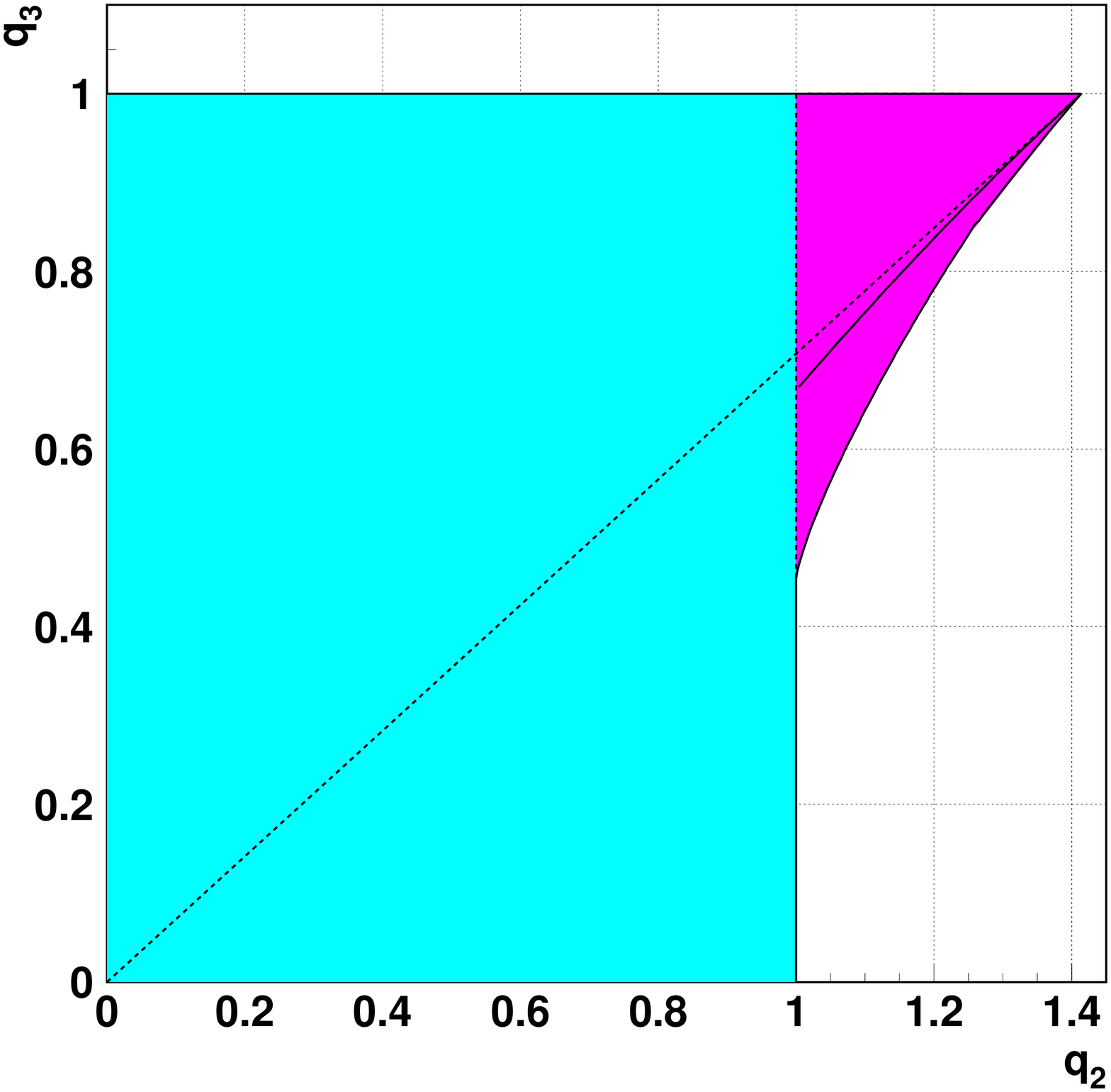}}
\setbox2=\vbox{\includegraphics*[width=8cm]{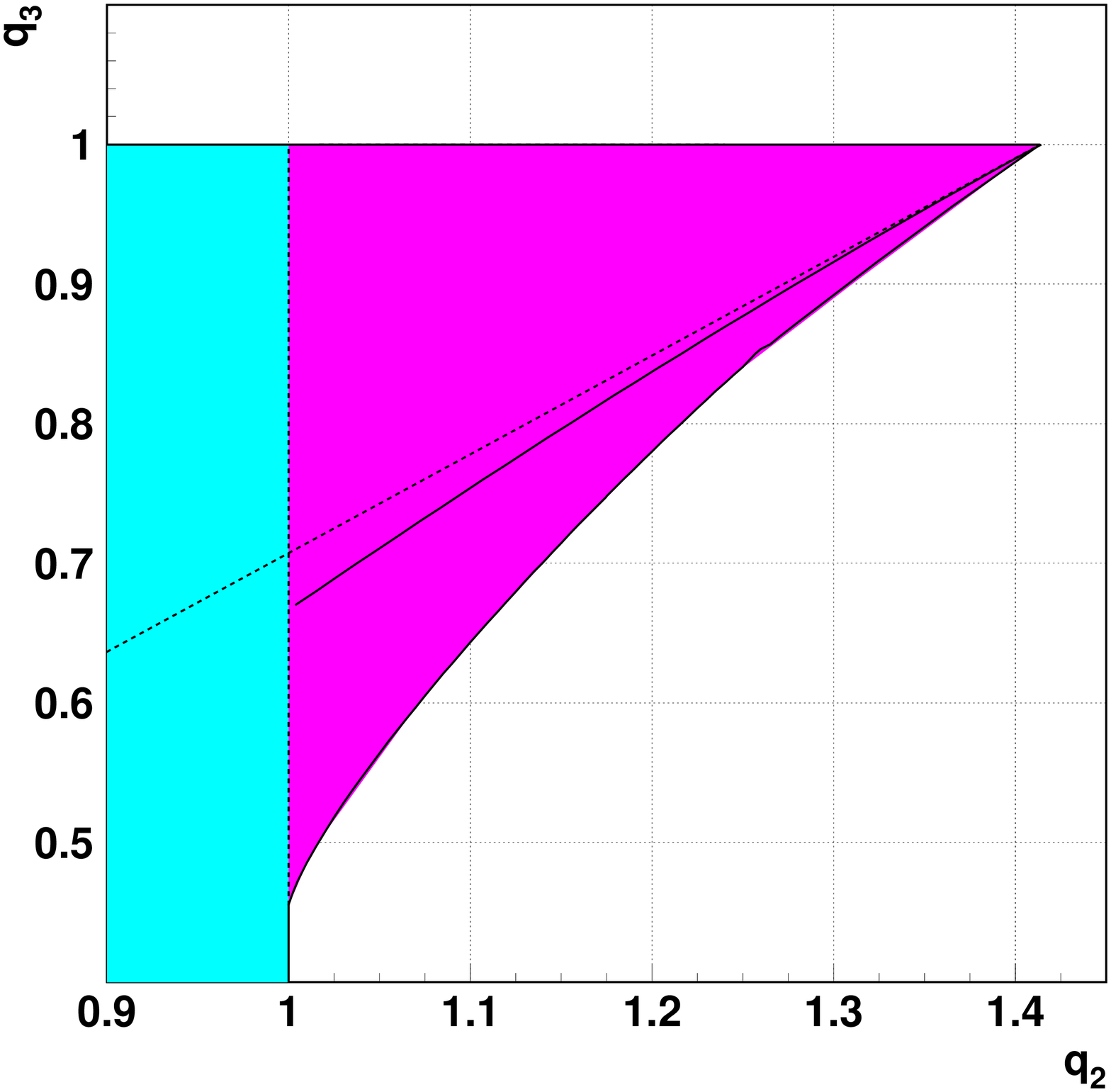}}
\centerline{\hglue 5.5cm\box1\hglue -4.5cm\box2}    
\caption{Same as Fig.~\protect\ref{Fig9}, for $m_{2}/m_{3}=2$. The 
dotted curve below (T) corresponds to the crude approximation of 
Eq.~(\protect\ref{approx-frontier}).}           
\label{Fig12}
\end{figure}
For $m_{2}=m_{3}$, in Fig.~\ref{Fig9}, we have a symmetric spike. 
The location of the peak at, $q_{2}=q_{3}=1.098$ reproduces fairly 
well the values given in the literature \cite{Baker90}.

For $m_{2}/m_{3}=1.1$, in Fig.~\ref{Fig10}, the spike leaves the 
horizontal line $q_{3}=1$ before $q_{2}=1$, as a generalized H$^-$ ion 
with unit charges and masses $(\infty,\, 1.1, \,1)$ is bound.  This is 
no longer the case for $m_{2}/m_{3}=1.5$, as seen in 
Fig.~\ref{Fig11}.

For $m_{2}/m_{3}=2$, shown in Fig.~\ref{Fig12}, the domain found in 
our variational calculation is flat.  In the lower part, it extends 
appreciably further than indicated by the crude approximation of 
Eq.~(\ref{approx-frontier}).

\section{Outlook}
\label{se:Outlook} 
Many questions remain open concerning  the stability of 3-charge systems.
Along the paper, we pointed out that in some limiting cases, more 
accurate results would be desirable. For instance, a question is 
whether very large values of the mass ratio $m_{3}/m_{2}$ exclude the 
possibility of binding with $q_{3}>1$.

There are also more general questions, concerning domains of some 
parameters for which stability will never be reached, whatever value 
is given to the other parameters.

For given masses $m_{i}$, the answer is immediate: there is always a 
set of charges, for instance $q_{3}<q_{2}<q_{1}=1$, that makes the 
system stable. 

For given charges $q_{2}$ and $q_{3}$, and $q_{1}=1$ to fix the scale, 
the situation is different: one has clearly three possibilities.  
Region \{1\} is the unit square $\{q_{2}<1,\, q_{3}<1\}$, where any mass 
configuration corresponds to a stable ion.  Region \{2\} includes for 
instance the point $q_{2}=2$ and $q_{3}=0.8$: there is sometimes 
stability, $m_{1}=\infty$, $m_{2}\ll m_{3}$ is an example, and 
sometimes breaking into an atom and a charge, as for $m_{1}=\infty$, 
$m_{2}\gg m_{3}$. Region \{3\} includes points like $q_{2}=q_{3}=2$ 
for which stability will never occur. Determining the properties of 
the boundary between regions \{2\} and \{3\} would be very interesting. 

A possible starting point is the  result by Lieb \cite{Lieb},
that for a fixed nucleus $\alpha_{1}=0$, $q_{1}=1$, a bound state will 
never occur if
\begin{equation}
\label{Lieb-cond}
{1\over q_2}+{1\over q_3}<1.
\end{equation}
A simple proof is given in Appendix A.  

This upper bound for possible stability at $m_{1}=\infty$ (i.e., 
stability occurring for at least some value of $m_{2}/m_{3}$) is not 
too far from the lower bound of Fig.~\ref{Fig13}, obtained from our 
variational method.  More extensive computations would be necessary to 
sketch the shape of the domain of absolute instability, in particular 
by relaxing the condition $m_{1}=\infty$.  Note that along the 
symmetry axis, the limit is $q_{2}=q_{3}\simeq 1.098$ for 
$m_{1}=\infty$ and $m_{2}=m_{3}$ finite, while it reaches 
$q_{2}=q_{3}\simeq 1.24$ for $m_{1}$ finite and $m_{2}=m_{3}=\infty$.
Thus, along the symmetry axis, the frontier 
between  regions \{2\} and \{3\} is saturated in the 
Born--Oppenheimer limit.  On the other hand, for $q_{2}\gg q_{3}$ or 
$q_{2}\ll q_{3}$, the question is whether this frontier 
has $q_{2}=1$ and $q_{3}=1$ as actual asymptotes, as tentatively 
pictured in Fig.\ref{Fig14} or reached these lines above some values of 
$q_{2}$ or $q_{3}$. 
\begin{figure}[hbct]
\setbox1=\vbox{\includegraphics*[width=8cm]{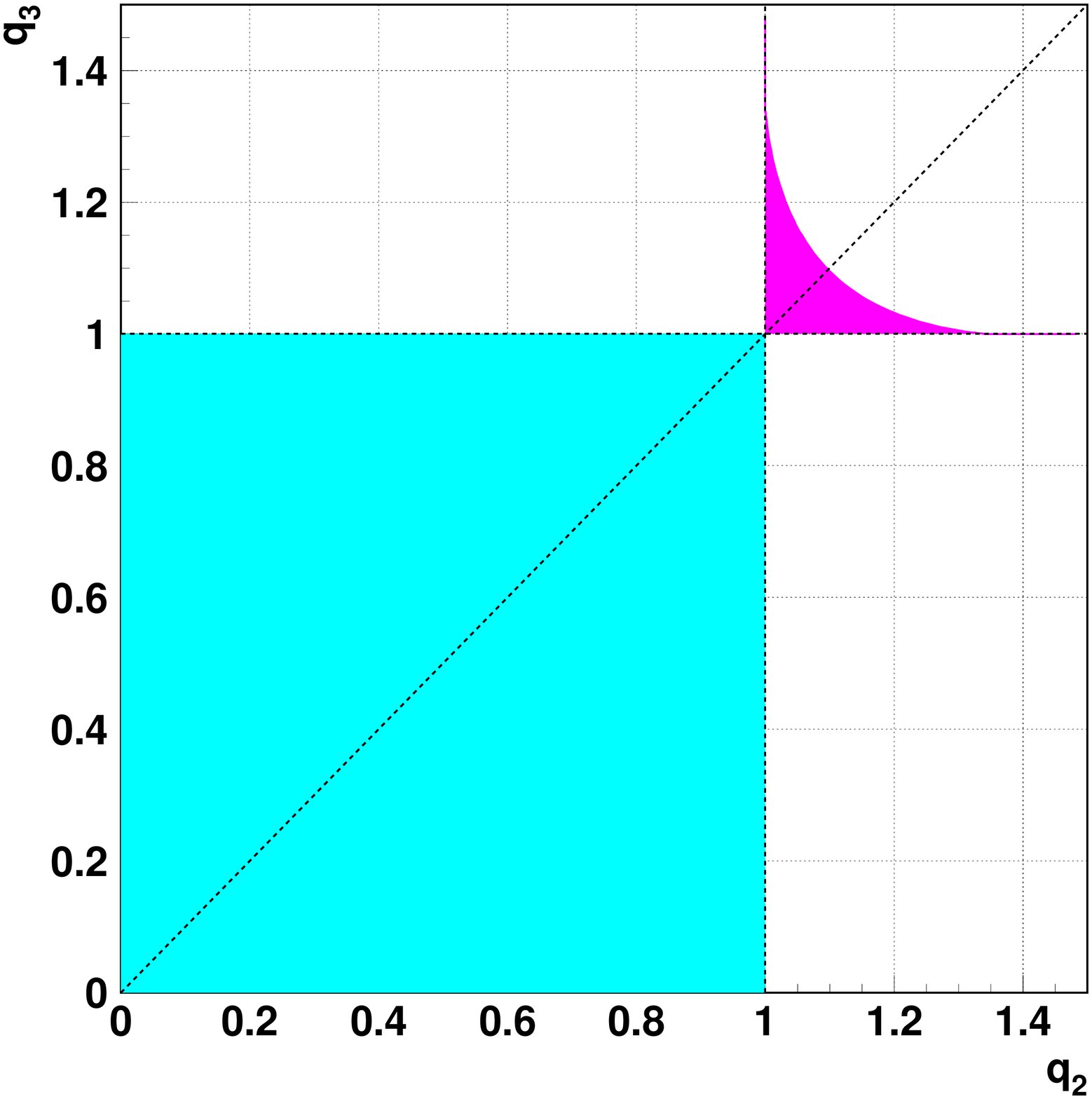}}
\setbox2=\vbox{\includegraphics*[width=8cm]{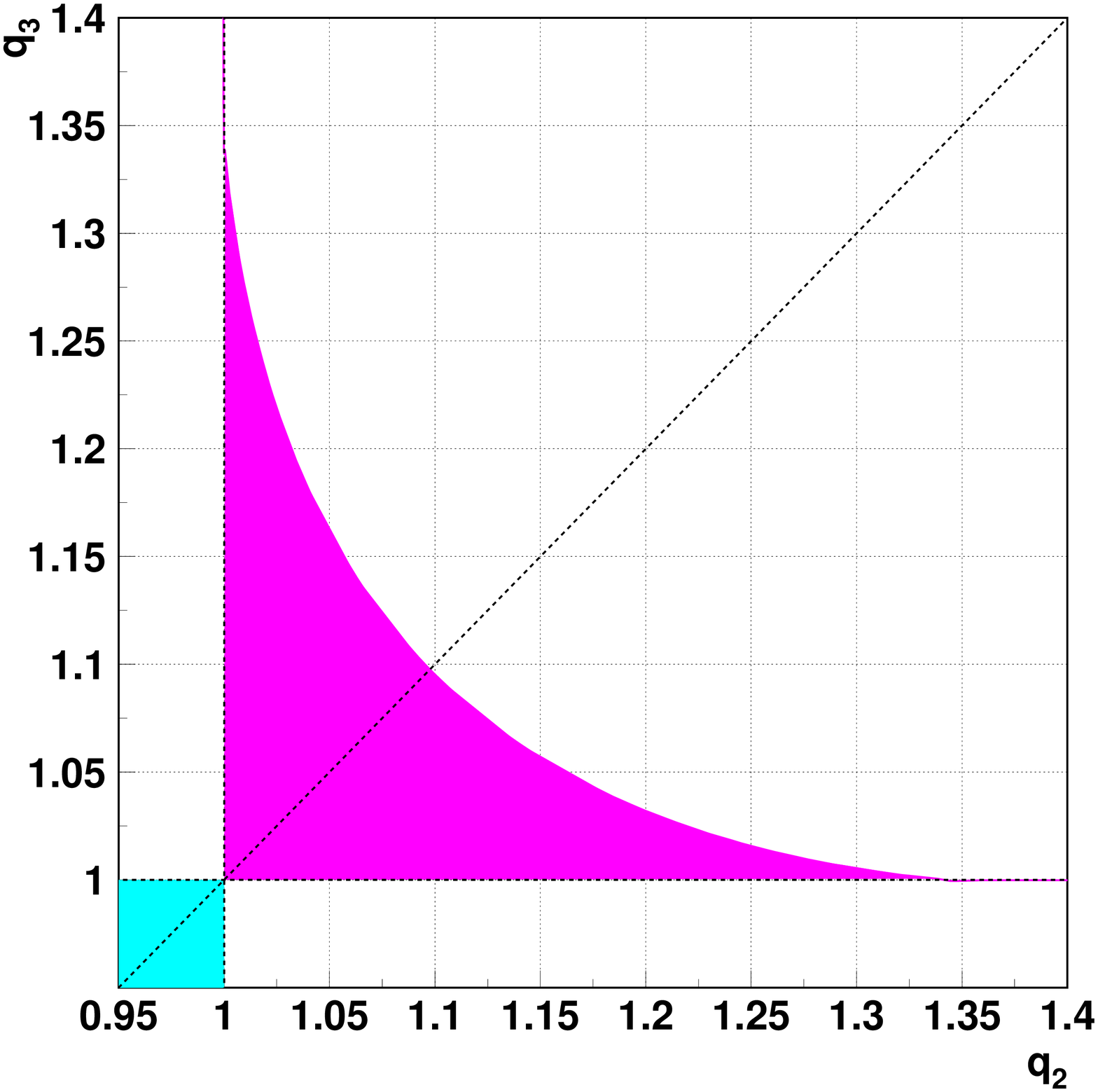}}
\centerline{\hglue 5.5cm\box1\hglue -4.5cm\box2}    
\caption{Variational estimate of the domain of possible stability 
beyond $q_{2}=1$ and $q_{3}=1$, for $m_{1}=\infty$. Inside the 
domain, stability occurs at least for some value of the mass ratio 
$m_{2}/m_{3}$.}           
\label{Fig13}
\end{figure}
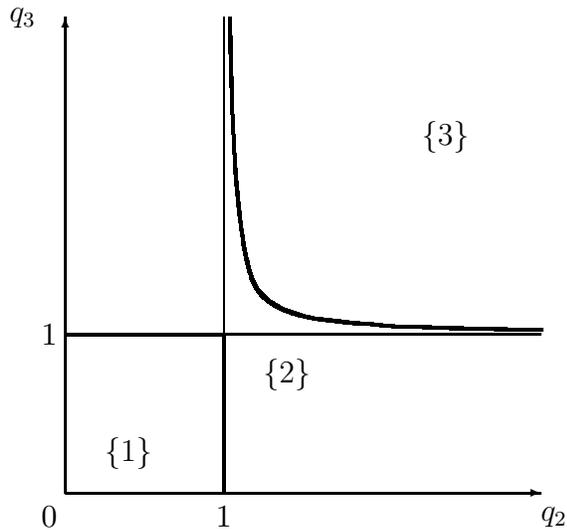
\begin{figure}[htbc]
  \vskip 1cm
 \begin{center}
 \setlength{\unitlength}{0.30pt}
  \begin{picture}(1000,700)(0,30)
  \font\gnuplot=cmr10 at 12pt
  \gnuplot
  \linethickness{.7pt}
  \put(300,100){\vector(1,0){600}}
  \put(300,100){\vector(0,1){600}}
  \linethickness{.4pt}
\put(500,100){\line(0,1){600}}
\put(300,300){\line(1,0){600}}
\put(270,70){\makebox(0,0)[l]{0}}
\put(900,70){\makebox(0,0)[l]{$q_{2}$}}
\put(230,700){\makebox(0,0)[l]{$q_{3}$}}
\put(350,150){\makebox(0,0)[l]{$\{1\}$}}
\put(550,250){\makebox(0,0)[l]{$\{2\}$}}
\put(750,550){\makebox(0,0)[l]{$\{3\}$}}
\put(500,70){\makebox(0,0)[c]{1}}
\put(270,300){\makebox(0,0)[l]{1}}
\linethickness{1.2pt}
\put(500,100){\line(0,1){200}}
\put(300,300){\line(1,0){200}}
\qbezier[150](506.,699.)(510.,398.)(541.,356.)
\qbezier[150](541.,356.)(554.,339.)(577.,330.)
\qbezier[150](577.,330.)(592.,324.)(613.,320.)
\qbezier[150](613.,320.)(628.,318.)(648.,316.)
\qbezier[150](648.,316.)(664.,314.)(684.,313.)
\qbezier[150](684.,313.)(700.,311.)(720.,310.)
\qbezier[150](720.,310.)(736.,310.)(755.,309.)
\qbezier[150](755.,309.)(772.,308.)(791.,308.)
\qbezier[150](791.,308.)(808.,307.)(827.,307.)
\qbezier[150](827.,307.)(844.,307.)(862.,306.)
\qbezier[150](862.,306.)(879.,306.)(898.,306.)
\end{picture}
  \end{center}
\vskip .3cm 
\caption{\label{Fig14} Guess at the shape of the border of the domain 
of absolute instability. In region \{1\}, binding occurs for any set 
of constituent masses. In region \{2\}, binding is achieved under 
some conditions for the masses. In region \{3\}, stability is never 
obtained.}
\end{figure} 
\section*{Acknowledgments}
One of us (T.T.W.) benefited from the warm atmosphere of the theory 
division at CERN.
%
%
\section*{Appendix A:\\ Proof of instability for 
\boldmath$q_{2}^{-1} + q_{3}^{-1}<1$\unboldmath}
\label{App1}
We give here a proof of the result on instability for all values of 
$m_{2}$ and $m_{3}$ if $q_{2}^{-1} + q_{3}^{-1}<1$, provided 
$\alpha_1=0$, with normalization $q_{1}=1$.

 First, it is shown that $r\vec{\rm p}^{2}$ is a 
positive operator, in the sense that any diagonal matrix element is 
positive. Indeed, separating the radial and angular part of 
$\vec{\rm p}^{2}$,
\begin{eqnarray}
	\langle\Psi\vert r\vec{\rm p}^{2}\vert\Psi\rangle 
& = & -\int r\Psi \Delta\Psi\, {\rm d}^{(3)}\vec{\rm r}
     \nonumber\\
& = & \int r\Psi {L^{2}\over r^{2}}\Psi\, {\rm d}^{(3)}\vec{\rm r}
      -\int {\rm d}\Omega\int \left[r\Psi {\partial^{2}(r\Psi)\over 
      \partial r^{2}}\right] r{\rm d}r 
      \nonumber\\
& = & \int r \left\vert \overrightarrow{\nabla}_{\Omega}\Psi\right\vert^{2}
       + \int {\rm d}\Omega\int  r{\rm d}r \left( {\partial (r\Psi)\over 
      \partial r}\right)^{2}
      \nonumber\\
 & = &\int {{\rm d}^{(3)}\vec{r}\over r}\left\vert\overrightarrow{\nabla} 
 (r\Psi)\right\vert^{2}.
	\label{eq:Lieb:1}
\end{eqnarray} 
Now consider the Hamiltonian
\begin{equation}
	H={\vec{\rm p}_{2}^{2}\over 2 m_{2}}+{\vec{\rm p}_{3}^{2}\over 2 m_{3}}
	-{q_{2}\over r_{2}}-{q_{3}\over r_{3}}+{q_{2}q_{3}\over r_{23}},
	\label{eq:Lieb:2}
\end{equation}
whose thresholds $(1,i)$ with $i=2,3$ are governed by the Hamiltonian
\begin{equation}
	h_{i}={\vec{\rm p}_{i}^{2}\over 2 m_{i}}
	-{q_{i}\over r_{i}}.
	\label{eq:Lieb:3}
\end{equation}
These $h_i$ and the 3-body Hamiltonian $H$ fulfill the identity
\begin{equation}
	\label{eq:Lieb:4}
{r_3}(H-h_2)+{r_2}(H-h_3)=
{r_2}{\vec{\rm p}_2^2\over 2m_2}+
{r_3}{\vec{\rm p}_3^2\over 2m_3}+
q_2q_3\left({r_2+r_3\over r_{23}}-{1\over q_2}-{1\over q_3}\right).
\end{equation}
In the r.h.s., the two first terms are always positive, and so is the third
one if $q_2^{-1}+q_3^{-1}<1$, due to the triangular inequality. Looking now
at the l.h.s., its expectation value is always positive, which means 
that 
\begin{equation}
\label{eq:Lieb:5}
\langle\Psi\vert r_{3}(H-h_{2})\vert\Psi\rangle>0
\quad\hbox{or}\quad
\langle\Psi\vert r_{2}(H-h_{3})\vert\Psi\rangle>0.
\end{equation}
In the first case take $\Psi$ as the ground state of $H$, which 
satisfies $H\Psi=E^{(3)}\Psi$. This translates into
\begin{equation}
\label{eq:Lieb:6}
E^{(3)}\langle\Psi\vert r_{3}\vert\Psi\rangle >
\langle\Psi\vert r_{3}\left({\vec{\rm p}_2^2\over 2m_2}-{q_{2}\over 
r_{2}}\right)\vert\Psi\rangle, 
\end{equation}
or
\begin{equation}
\label{eq:Lieb:7}
E^{(3)}\langle\sqrt r_{3}\Psi\vert\sqrt r_{3}\Psi\rangle >
\langle\sqrt r_{3}\Psi\vert \left({\vec{\rm p}_2^2\over 2m_2}-{q_{2}\over 
r_{2}}\right)\vert\sqrt r_{3}\Psi\rangle\ge
E^{(2)}_{12}\langle\sqrt r_{3}\Psi\vert\sqrt r_{3}\Psi\rangle   
\end{equation}
from the variational principle. In the second case 
\begin{equation}
\label{eq:Lieb:8}
E^{(3)}>E^{(2)}_{13}  
\end{equation}
So, if $m_{1}=\infty$ and $q_{2}q_{3}>q_{2}+q_{3}$, either 
$E^{(3)}>E^{(2)}_{12}$ or $E^{(3)}>E^{(2)}_{13}$.
\section*{Appendix B: Variational method}
\label{App:Varia}
We briefly describe the variational method used for the numerical 
results displayed in this paper.  More details can be found in 
\cite{Krikeb98}.  The ground state of the Hamiltonian 
(\ref{Hamiltonian}) has been searched using trial wave functions of 
the type \cite{Chandrasekhar44} %
\begin{equation}
\label{var-w-f}
\Psi=\sum_{i}C_{i}\varphi_{i}=
\sum_{i}C_{i}\left[\exp(-a_{i} r_{23} - b_{i} r_{31} - c_{i} r_{12} 
)+\cdots\right]
\end{equation}
from which all matrix elements can be calculated in close form.  The 
dots are meant for similar terms obtained by permutation, in the case of 
identical particles.  For given range parameters, the weights $C_{i}$  
are listed in  a vector $\mathbf{C}$, which is found, together with the 
variational energy $\epsilon$, from the matrix equation 
\begin{equation}
\label{mat-eq}
\left(\widetilde{T}+\widetilde{V}\right) 
\mathbf{C}=\epsilon\widetilde{N}\mathbf{C},
\end{equation}
involving the restrictions of the kinetic and potential energy to the 
space spanned by the $\varphi_{i}$, whose scalar products are stored 
in the positive-definite matrix $\widetilde{N}$. 

As the number of terms increases, it quickly becomes impossible to 
determine the best range parameters, even with powerful minimization 
programs, as too many neighboring sets give comparable energies.  
One way out \cite{Kamimura88} consists of imposing all $a_{i},\, b_{i}$ and $c_{i}$ to 
be taken in a geometric series. Then only the smallest and the largest 
have to be determined numerically. For instance, this method allows 
one to reproduce the binding energy $-0.262005$ of the Ps$^-$ ion, in 
agreement with the best results in the literature.

The question now is to find the frontier. Let us consider, for instance, 
the problem of Sec.~\ref{Upper-part}. Here $m_{1}=\infty$, $q_{1}=m_{2}=1$, 
and $q_{2}/q_{3}=m_{3}^{1/2}$ when one searches the limit of 
stability among the threshold separation (T).

One can estimate the ground-state energy  of
\begin{equation}
{ \vec{\rm p}_{2}^{2}\over 2} +  { \vec{\rm p}_{3}^{2}\over 2 m_{3} }
-{m_{3}^{1/2} q_{3}\over r_{2} } -{ q_{3}\over r_{3} } 
+ {m_{3}^{1/2} q_{3}^{2}\over r_{23} },
\end{equation}
starting from some low value of $q_{3}$, and examine, by suitable 
interpolation, for which $q_{3}$ it matches the threshold 
$E_{\rm the}=-q_{3}^{2} m_{3}/2$.

A more direct strategy consists to set $E=E_{\rm the}$ in the 
Schr{\"o}dinger equation, apply  a rescaling and solve
\begin{equation}
{ \vec{\rm p}_{2}^{2}\over 2} +  { \vec{\rm p}_{3}^{2}\over 2 m_{3} }
-{m_{3}^{1/2} \over r_{2} } -{ 1\over r_{3} }+{m_{3}\over 2}= 
- q_{3}\, {m_{3}^{1/2} \over r_{23} },
\end{equation}
using the same trial function (\ref{var-w-f}), resulting in a matrix 
equation very similar to (\ref{mat-eq}), where the positive-definite 
matrix $\widetilde{N}$ now represents the restriction of 
$m_{3}^{1/2}/r_{23}$ in the space of the $\varphi_{i}$.  In principle, the 
variational wave function needs not to be normalizable at threshold, 
but this becomes immaterial as soon as very long range components are 
introduced in the expansion (\ref{var-w-f}).  It was checked that the 
extrapolation method and the direct computation give  the same 
result for the frontier.
%

%

\end{document}